\documentclass[12pt,a4paper,twoside]{article}

\usepackage{epsfig}
\usepackage{graphics}
\usepackage{multicol}
\usepackage[dvips]{lscape}
\usepackage{amsmath,amsthm,amsfonts,amssymb}

\begin{document}

\def\thefootnote{\fnsymbol{footnote}}

\thispagestyle{empty}
\markboth{{\footnotesize \rm G.~Mavromanolakis}}
{{\footnotesize \bf Quartz fiber calorimetry and calorimeters}}
\pagestyle{myheadings}


\vspace*{-45pt}
\hfill {\tt HEP-Cavendish/04/34}

\vspace*{-10pt}
\hfill {\tt UA-NPPS/06/2003}


\vspace*{25pt}

{\LARGE\bf Quartz fiber calorimetry and calorimeters
\footnote[1]
{~Material based on notes and seminars given at 
University of Athens, Department of Physics}
}
\label{review_qcal}
%

\vspace{25pt}
G.~Mavromanolakis~\footnote[7]
{
~email: {\tt gmavroma@hep.phy.cam.ac.uk} or {\tt gmavroma@mail.cern.ch}
}

{\em \small
University of Cambridge, Department of Physics\\
Cavendish Laboratory, High Energy Physics Group\\
Madingley Road, Cambridge, CB3 0HE, UK
}


\vspace{25pt} 
{\small
Quartz fiber calorimetry is a technique the signal generation mechanism of 
which is based on the Cherenkov effect. In this article we try to give a 
comprehensive overview of the subject. We start with a general introduction to 
calorimetry where the basic elements that characterize the development of 
electromagnetic and hadronic showers are discussed. Then we describe in detail 
the operation principle and the properties of calorimeters equipped with 
quartz fibers. The main advantages of this type of calorimeters are the 
radiation hardness, the fast response and the compact detector dimensions, 
features that derive from the quartz material and the specific mechanism of 
operation. A section is devoted to presenting the quartz fiber calorimeters 
that have been built or planned to in various experiments to operate as 
centrality detectors, trigger detectors, luminosity monitors or general 
purpose very forward calorimeters.
}

{\small
\tableofcontents
}

\newpage

\section{Introduction to calorimetry}

Calorimetry comprises the experimental methods one develops to perform
energy measurement. A calorimeter~\cite{ref:QC:cushman92}-\cite{ref:QC:wigmans2000} 
is a detector, literally a block of matter with proper instrumentation, which 
measures the energy of incident particles. Its characteristic feature is that 
the resolution of the energy measurement improves with energy, while the size 
of the detector scales logarithmically with it. Also of importance is the fact 
that a calorimeter is a detector which is sensitive to both charged and 
neutral incident particles. Depending on appropriate construction, 
calorimeters can also provide position measurement and particle 
identification in addition to energy measurement. The signal generation 
mechanism in a calorimeter is as follows: the incident particle interacts 
inside the calorimeter volume and initiates a shower of secondary particles, 
the shower develops and its products generate signal by passing through 
the sensitive material of the calorimeter.
The operation principle of the majority of calorimeters is based on the 
$dE/dx$ technique, i.e. the signal depends on the energy deposited by the 
shower particles in the sensitive material, and is generated by scintillation
or ionization.
The signal can be light, for calorimeters with scintillators, or ionization 
charge, for those with gaseous or semiconductive sensitive material. 
Calorimeters with a different principle of operation are those composed of 
lead glass or quartz fibers. For these, the signal generation mechanism is 
based on the Cherenkov effect.

Calorimeters can be distinguished between electromagnetic and hadronic, 
whether they are designed to measure the energy of incident electrons, 
positrons and photons, or hadrons, respectively. They are also categorized in
homogeneous and sampling calorimeters, depending on their construction. In a 
homogeneous calorimeter, the whole volume is considered sensitive, i.e. shower
production and development and signal generation and collection occur at the 
same material. The main advantage of this type of calorimeters is their excellent
performance with respect to energy resolution. A sampling calorimeter is 
composed of two different materials. A passive material, the absorber, where 
the shower develops, and the sensitive or active material where the signal is
produced and collected. The absorber is usually copper, steel, lead or 
other dense metal. The sensitive material, that can be scintillator plates, 
gas tubes or silicon layers, is placed between consecutive absorber layers 
(``sandwich'' structure) or is uniformly distributed inside the absorber 
volume, as in the case of scintillation tiles or fibers (``spaghetti'' 
structure). Sampling calorimeters have worse energy resolution because the 
statistical fluctuations in production and collection of their signal are 
significantly larger compared to homogeneous ones. However, they cost less, 
have compact size and are the only practical means one can use to measure the 
energy of hadrons.

\subsection{Electromagnetic shower development}

\begin{figure}[tb]
  \centering
  \begin{tabular}{cc}
    \epsfxsize=204pt
    \epsffile{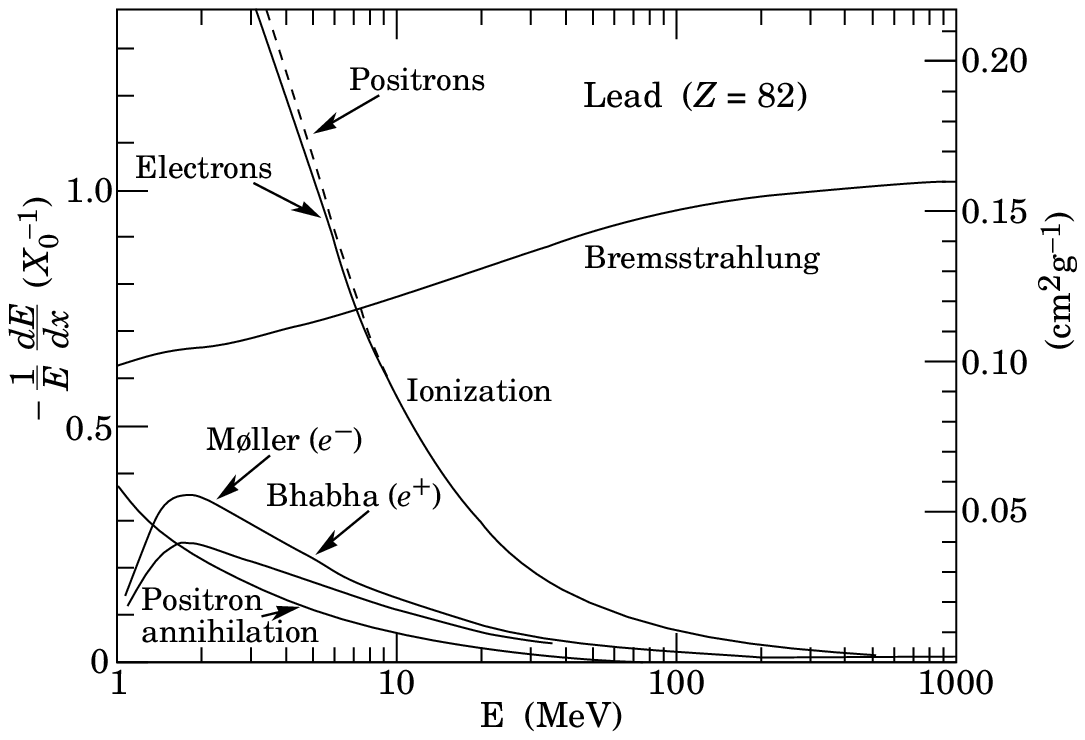}
    &
    \epsfxsize=180pt
    \epsffile{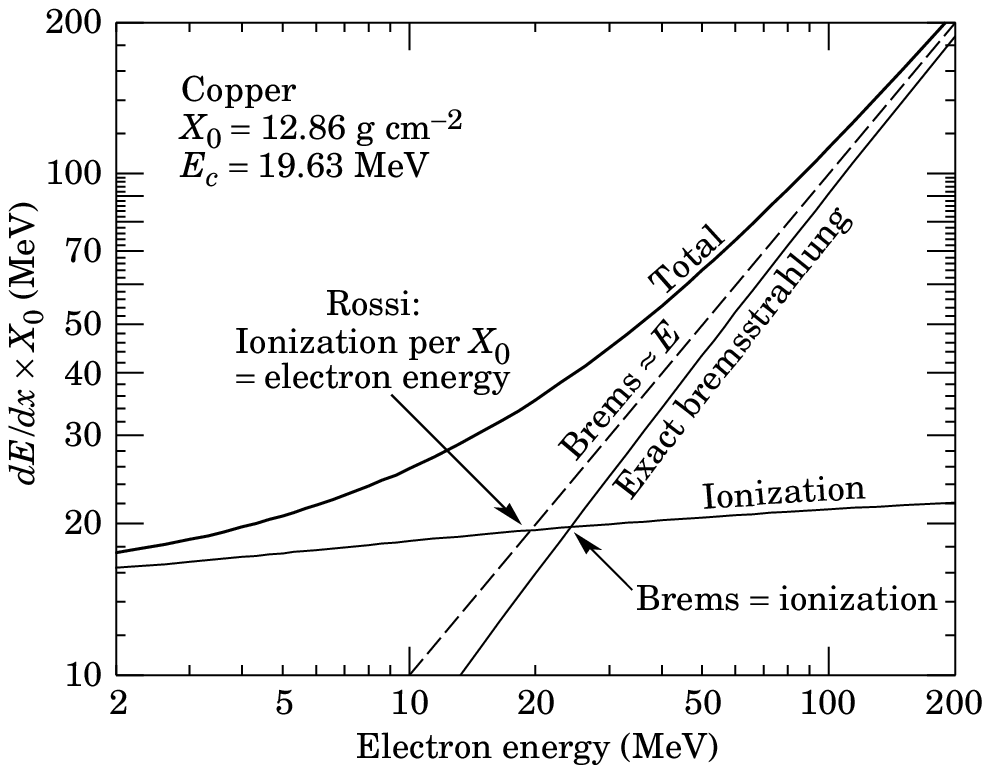}
    \\
    (a)  & (b)  \\
  \end{tabular}
  \caption{(a) relative energy loss per unit length 
  ($\frac{1}{E}\frac{dE}{dx}$) in lead for different processes as a function 
  of electron or positron energy, (b) energy loss per unit length from 
  ionization and bremsstrahlung as a function of electron energy (in copper).
  Both forms of defining $E_c$ are shown, according to 
  $\left(\frac{dE}{dx}\right)_{ion}=\left(\frac{dE}{dx}\right)_{brem}$
  or $\left(\frac{dE}{dx}\right)_{ion}=\frac{E_c}{X_0}$.}
  \label{fig:QC:elossfrac+encrit_cu}
\end{figure}

For energies above 1~GeV, electrons and positrons lose energy through
bremsstrahlung, and photons through pair production
\cite{ref:QC:cushman92}-\cite{ref:QC:pdg2000}.
The secondary particles produced interact through the same processes and a 
shower develops inside the calorimeter. For lower energies 
(E $\lesssim$ 10~MeV) electrons and positrons interact with matter and lose 
energy mainly through ionization of the atoms of the detector. Additional 
processes such as bremsstrahlung, Moller scattering, Bhabha scattering or 
positron annihilation do not contribute significantly to energy loss, as 
shown in fig.~\ref{fig:QC:elossfrac+encrit_cu}(a) \cite{ref:QC:pdg2000}.
The shower development ceases when energy loss from ionization is greater 
than that from bremsstrahlung. The energy loss per unit length from 
ionization is practically constant, since it depends logarithmically on 
electron or positron energy, $|dE/dx|_{ion}\propto \ln E$, in contrast to 
energy loss from bremsstrahlung which shows linear dependence on energy, 
$|dE/dx|_{brem} \propto E $ 
(fig.~\ref{fig:QC:elossfrac+encrit_cu}(b) \cite{ref:QC:pdg2000}).
We define {\em critical energy}, $E_c$, as the energy at which the loss rates
from ionization and from bremsstrahlung become equal. The critical energy is 
a crucial parameter which characterizes the material of a calorimeter. It is 
well approximated by $E_c \approx 800/(Z + 1.2)$. Better representation of 
experimental results, as far as transverse shower development is concerned, 
can be achieved by using the following approximation \cite{ref:QC:pdg2000}
\begin{eqnarray}
E_c = \frac{610}{Z + 1.24} \ \textrm{MeV}, & \textrm{for liquids or solids}
\end{eqnarray}
\begin{eqnarray}
E_c = \frac{710}{Z + 0.92} \ \textrm{MeV}, & \textrm{for gases}
\end{eqnarray}
where $Z$ is the atomic number of material.

The {\em radiation length}, $X_0$, is another characteristic quantity, which 
is related to the longitudinal development of electromagnetic showers. It is 
defined as the mean distance that an electron traverses to lose $(1-1/e)$ of 
its energy by bremsstrahlung. It is approximately 
$X_0 \approx 180 \cdot A/Z^2$ (in gr/cm$^2$), where $A$, $Z$ are mass and 
atomic numbers of material, respectively,
\begin{eqnarray}
X_0 = \frac{716.4 \cdot A}{Z(Z+1) \cdot \ln(287/\sqrt{Z})} \ \textrm{gr/cm$^2$}
\end{eqnarray}
For a photon, the mean distance that it travels before it converts to a 
$e^- e^+$ pair with probability $(1-1/e)$ is equal to $\frac{9}{7} X_0$.

The mean longitudinal profile of the energy deposition of a shower, or 
equivalently the longitudinal distribution of signal production, can be 
described by a $\Gamma$-distribution function 
\cite{ref:QC:cushman92,ref:QC:amaldi81,ref:QC:pdg2000},
\begin{eqnarray}
\label{eq:QC:em_longitudinal_profile}
 \frac{dE}{dt} = E \cdot b \cdot \frac{(bt)^{\alpha -1} e^{-bt}}{\Gamma(\alpha)}
\end{eqnarray}
where $t = x/X_0$, $y = E/E_c$ and $b \approx 0.5$, $\alpha -1 = b t_{max}$, 
with $t_{max}$ the point where the distribution has its maximum,
\begin{eqnarray}
 t_{max} = 1.0 \cdot (\ln y + C)
\end{eqnarray}
with values commonly used, $C$=-0.5 or -1.0 for electrons, $C$=+0.5 or -0.5 
for photons. The energy of the electromagnetic shower is contained at 95\% 
level within distance $L_{EM}(95\%)$, which is approximated by
\begin{eqnarray}
 L_{EM}(95\%) = X_0 \cdot (t_{max} + 0.08 Z + 9.6)
\end{eqnarray}

The shower develops also in the transverse direction, due to multiple 
scattering of electrons by nuclei's field (multiple Coulomb scattering). The 
characteristic length that governs the transverse development of an 
electromagnetic shower is the so called {\em Moli$\grave{{e}}$re radius}, 
$R_M$. It is defined as 
\begin{eqnarray}
 R_{M} = X_0 \cdot E_s/E_c
\end{eqnarray}
where $E_s = \sqrt{4\pi/\alpha} m_e c^2 = 21.2$~MeV. A simpler approximate 
expression is $R_M \approx 7 \cdot A/Z$~(in gr/cm$^2$). In general, 90\% 
of shower energy is contained within a cylinder of radius 
$R_{EM}(90\%)=R_M $, 95\%(99\%) in a radius of $2~R_M (3.5~R_M)$ respectively.

When the absorber of a calorimeter is a mixture of different elements, the 
effective radiation length $X_0$ and Moliere radius $R_M$ can 
be calculated with
\begin{eqnarray}
 1/X_0 = \sum w_i/X_{0i} & \textrm{and} & 1/R_M = \frac{1}{E_s} \sum \frac{w_i E_{c i}}{X_{0 i}}
\end{eqnarray}
where $w_i$ is the composition by weight and $X_{0i}$, $E_{c i}$, the 
radiation length and the critical energy of the corresponding element.

The total amount of energy ($E_{visible}$) responsible for signal production
(scintillation light or charge) is proportional to the total energy loss
of shower particles through ionization. Practically $|dE/dx|_{ion}$ is constant 
with energy. Thus, the signal that is produced in a calorimeter is 
proportional to the total track length ($T$) traveled by the charged particles of the 
shower, hence it is
\begin{eqnarray}
\label{eq:QC:E_vis}
 E_{visible} \propto E_{ionization} = T \cdot |dE/dx|_{ion} \propto T
\end{eqnarray}
Assuming that on average, an electron or a positron interacts through 
bremsstrahlung after traversing length $X_0$, and that a photon converts to 
a $e^- e^+$ pair per $X_0$, then after traveling distance $t=x/X_0$ inside 
the calorimeter absorber, the shower will consist of $N=2^t$ particles, with 
$E/N$ energy per particle. Production of new particles stops typically when 
almost every particle in the shower carries energy around $E_c$. Therefore 
the total particle multiplicity of the shower is roughly equal to $E/E_c$, 
and every particle has traversed distance $X_0$ on average. Thus, the total 
track length inside the calorimeter is
\begin{eqnarray}
\label{eq:QC:total_track}
 T = X_0 \cdot E/E_c
\end{eqnarray}
Equations \ref{eq:QC:E_vis} and \ref{eq:QC:total_track} clearly manifest 
the fundamental operation principle of a calorimeter: the produced signal is 
proportional to the energy of the incident particle.

\subsection{Hadronic shower development}

The incidence of a high energy hadron in a 
calorimeter~\cite{ref:QC:wigmans91,ref:QC:wigmans87},\cite{ref:QC:cushman92}-\cite{ref:QC:amaldi81}
produces a shower of particles due to inelastic collisions with nucleons of 
the absorber's nuclei. Secondary particles are produced, mainly pions and 
nucleons, in multiplicity which increases logarithmically with energy per 
collision. On average 1/3 of the produced pions are $\pi^0$'s, that 
subsequently decay into photons and generate electromagnetic showers. The 
fraction of the energy of hadrons that does not dissipate in particle 
production is lost through interactions of excitation of nuclei. Generally 
this type of interaction does not contribute to signal production.

The characteristic length which governs the longitudinal development of 
hadronic showers is the {\em interaction length}, $\lambda_I$. In a way 
similar to the radiation length for electromagnetic showers, it is defined as 
the mean distance that a hadron traverses to lose $(1-1/e)$ of its energy in 
inelastic collisions. It is given by
\begin{eqnarray}
\lambda_I = \frac{A}{N_{Av} \cdot \rho \cdot \sigma_{inel}}~\textrm{cm} \approx 35 \cdot A^{1/3}~\textrm{gr/cm$^2$}
\end{eqnarray}
where $A$, $\rho$ is the mass and the density of the material, respectively, 
$N_{Av}$ is the Avogadro number and $\sigma_{inel}$ is the cross section of 
inelastic interaction for proton.

The longitudinal profile of the energy deposition of a hadronic shower can be 
parameterized by a sum of a function with the form shown in 
eq.~\ref{eq:QC:em_longitudinal_profile} and a descending exponential, 
representing  the purely electromagnetic and the purely hadronic component of 
the shower, respectively. From experimental data the maximum of the shower 
occurs at depth $t_{max}$, which is approximately
\begin{eqnarray}
 t_{max} \sim 0.2 \ln E + 0.7
\end{eqnarray}
with $t_{max}$ in units of $\lambda_I$ and $E$ in GeV. The total depth that 
is sufficient for 95\% containment of the shower energy is
\begin{eqnarray}
 L_{had}(95\%) \simeq t_{max} + 2.5 \cdot \lambda_I \cdot E^{0.13}, &  (\textrm{$E$ in GeV})
\end{eqnarray}

The transverse development of a hadronic shower is determined by the mean 
transverse momentum of the produced particles, which is roughly 
$\langle p_t \rangle \simeq$~0.35~GeV. The transverse development does not 
scale with $\lambda_I$, however 95\% of the shower  energy is contained in a 
cylinder of radius $R_{had}(95\%) \lesssim 1 \cdot \lambda_I$.

The signal that is generated by a hadronic shower is lower than that produced by
an electromagnetic one of same energy. This is because a significant fraction,
about 25\%, 30\%, of the total energy dissipated by the hadronic shower inside the 
calorimeter is lost in nuclear break-up and excitation processes that do not 
contribute to detectable signal. This intrinsic difference in the response to the 
two different types of showers is expressed by the {\em degree of compensation} or 
{\em ratio $e/h$}. This quantity is a characteristic parameter of a 
calorimeter and is determined by its operation principle and its construction.
The $e/h$ ratio does not depend on energy, it can be derived indirectly by 
the $e/\pi$ ratio which is energy dependent. The $e/\pi$ is the ratio of the 
signal of an incident electron to the signal produced by the shower of an 
incident charged pion of same energy. The relation between the two ratios is 
given by
\begin{eqnarray}
\label{eq:QC:epi_eh}
\frac{e}{\pi}(E) = \frac{e/h}{1+(e/h-1) \cdot f_{em}(E)}
\end{eqnarray}
where $f_{em}(E)$ is defined as the average fraction of energy which contributes to 
the electromagnetic component of a hadron shower, mainly through $\pi^{0}$ 
production. The $f_{em}$ function has been studied theoretically and can be 
approximated by the expressions
\cite{ref:QC:acosta92,ref:QC:wigmans88,ref:QC:ganel95,ref:QC:gabriel94,ref:QC:groom97}
\begin{eqnarray}
\label{eq:QC:fem_groom}
f_{em}(E)=1-\left(\frac{E}{E_{o}}\right)^{m-1}
%
& \textrm{or} &
%
f_{em}(E)=k\cdot ln\left(\frac{E}{E_{o}}\right)
\end{eqnarray}
where $E_{o}$, $m$, $k$ are free parameters in the range $E_{o}$:~0.7-1~GeV, 
$m$:~0.8-0.9, $k$:~0.1-0.15.

\subsection{Energy resolution of calorimetric measurement}
\label{subsec:QC:energy_resolution}

The development of a shower and the production and collection of its signal 
are processes with statistical fluctuations, the contributions of which 
determine the energy resolution of a calorimeter. The number of sources and 
the amplitude of fluctuations are larger for hadronic showers than for 
electromagnetic ones, therefore electromagnetic calorimeters show better 
energy resolution than hadronic ones.

{\bf Energy resolution for electromagnetic shower}

In eq.~\ref{eq:QC:total_track} the total track length of an electromagnetic 
shower is estimated. We expect the produced signal to be proportional to the 
total track length. Practically in a calorimeter, only particles with energy 
greater than a threshold value $E_{cut}$ ($\sim$ 0.5~MeV) generate detectable 
signal, thus it is more meaningful to use the total detected track length, 
$T_d$. It is
\begin{eqnarray}
\label{eq:QC:total_track_detected}
T_d = F(\zeta) \cdot T = F(\zeta) \cdot X_0 \frac{E}{E_c}
\end{eqnarray}
where the fraction $F(\zeta)$ is approximated by the formula 
\cite{ref:QC:amaldi81}
\begin{eqnarray}
F(\zeta) \simeq e^{\zeta} \cdot \left( 1 + \zeta \ln\left(\frac{\zeta}{1.526}\right)\right)
%
&
\textrm{with}
&
\zeta = 4.58 \frac{ZE_{cut}}{AE_c}
\end{eqnarray}
In an ideal homogeneous electromagnetic calorimeter of infinite size, without 
energy leakage, the only source of fluctuations in signal production is the 
statistical fluctuation of length $T_d$, i.e. $\sigma(E)/E=\sigma(T_d)/T_d$. 
This is called {\em intrinsic fluctuation} because it describes the 
fluctuation which is related with the fundamental processes that govern the 
production and development of an electromagnetic shower. Its contribution to 
the energy resolution is about 0.7\%/$\sqrt{E(\textrm{GeV})}$.
\begin{eqnarray}
\left(\frac{\sigma(E)}{E}\right)_{intrinsic} \simeq \frac{0.7\%}{\sqrt{E(\textrm{GeV})}}
\end{eqnarray}
If a fraction of the shower energy leaks out of the calorimeter, then it is
\begin{eqnarray}
\label{eq:QC:fluctuations_leakage}
\left(\frac{\sigma(E)}{E}\right)_f \simeq \left(\frac{\sigma(E)}{E}\right)_{f=0} \cdot (1+4f)
\end{eqnarray}
where $f$ is the estimated fraction of energy leakage.

For calorimeters which detect light from scintillation or Cherenkov radiation,
statistical fluctuations on the production and collection of photons
({\em photostatistics}) contribute to the total energy resolution. For low 
light yield this fluctuation is significant, because the mean number of 
photoelectrons at photomultipliers per GeV of deposited energy in the 
calorimeter ($n_{pe}$) is small. It holds $\sigma_{pe} = \sqrt{n_{pe}}$, and 
thus it is
\begin{eqnarray}
\label{eq:QC:photostatistics}
\left(\frac{\sigma(E)}{E}\right)_{photostatistics} = \sqrt{\frac{1}{n_{pe}}} \cdot \frac{1}{\sqrt{E(\textrm{GeV})}}
\end{eqnarray}

In sampling calorimeters, in addition to these sources of fluctuations, 
we also have {\em sampling fluctuations}. These represent the statistical 
fluctuations in the number of $e^-$'s and $e^+$'s that traverse the sensitive 
material between the absorber layers. If $d$ is the thickness of each absorber 
layer and $T$ the total track length (eq.~\ref{eq:QC:total_track}), then the 
number of tracks that traverse the sensitive material is equal to $T/d$. 
Using instead the total detectable track length, $T_d$ 
(eq.~\ref{eq:QC:total_track_detected}), the energy resolution reads
\begin{eqnarray}
\left(\frac{\sigma(E)}{E}\right)_{sampling} \simeq \frac{1}{\sqrt{T_d/d}} =
\frac{\sqrt{E_c/F(\zeta)} \cdot \sqrt{d/X_0}}{\sqrt{E}}
\end{eqnarray}
Using the approximation $E_c \simeq \frac{550}{Z}$~MeV \cite{ref:QC:amaldi81},
which has a relative error less than $\pm 10\%$
for $13\leq Z \leq 92$, and the correction on $d$ by a factor of 
1/$\cos\theta$, with 
$\langle\cos\theta\rangle \simeq \cos\left(\frac{E_s}{\pi E_c} \right)$
to account for the mean angle of the tracks of the shower particles with respect to the 
direction of the incident particle, we have
\begin{eqnarray}
\left(\frac{\sigma(E)}{E}\right)_{sampling} \simeq
 3.2\% \sqrt{\frac{550}{Z F(\zeta) \cos\left(\frac{E_s}{\pi E_c} \right)}} \cdot \frac{\sqrt{d/X_0}}{\sqrt{E(\textrm{GeV})}}
\end{eqnarray}
or equivalently 
$\frac{\sigma(E)}{E} \simeq R_{EM} \cdot \frac{\sqrt{d/X_0}}{\sqrt{E(\textrm{GeV})}}$
with $R_{EM}$ about equal to 15-20\%.

There are also contributions to sampling fluctuations from
Landau fluctuations and path length fluctuations \cite{ref:QC:amaldi81}. 
They are negligible for calorimeters with liquid or solid 
sensitive material, whereas for gaseous calorimeters we should take them 
into account as they could contribute up to 5\% to 10\% per $\sqrt{E}$.

{\bf Energy resolution for hadronic shower}

The intrinsic fluctuations for hadronic showers are large and have a 
significant contribution to the total energy resolution. This is due to the 
fact that on average about 1/4 of the energy that is deposited in the 
calorimeter is dissipated through processes that do not contribute to 
generation of visible signal. This practically invisible fraction is 
dominated by large fluctuations, and consequently, the remaining detectable 
amount of energy shows large fluctuations as well. Besides, the purely 
electromagnetic component of a hadronic shower is generated mainly by the 
prompt $\pi^0$'s that are produced in the initial interactions. And since 
the number of $\pi^0$'s is small, in approximate it increases logarithmically 
with energy, it has large statistical fluctuations contributing significantly 
to the total signal fluctuation. In general, no matter what the calorimeter 
structure and its operation principle are, the intrinsic fluctuation 
contribution to the energy resolution for hadronic showers is roughly
\begin{eqnarray}
\left(\frac{\sigma(E)}{E}\right)_{intrinsic} \simeq \frac{45\%}{\sqrt{E(\textrm{GeV})}}
\end{eqnarray}

Contributions from energy leakage and photostatistics are approximated by 
equations \ref{eq:QC:fluctuations_leakage} and \ref{eq:QC:photostatistics}, 
as in the case of electromagnetic showers.

Sampling fluctuations are expected to have similar dependence on absorber layer 
thickness and on energy as for the electromagnetic showers,
\begin{eqnarray}
\left(\frac{\sigma(E)}{E}\right)_{sampling} \simeq
 R_{had} \cdot \frac{\sqrt{d/X_0}}{\sqrt{\frac{3}{4} \cdot E(\textrm{GeV})}}
\end{eqnarray}
with $R_{had}$ about 30\% to 40\% or roughly $R_{had}\sim 1.5 \ \textrm{to} \ 2 \cdot R_{EM}$
\cite{ref:QC:amaldi81}.
The fraction 3/4 expresses the average percentage of the total energy which 
is deposited in calorimeter through ionization by electrons, positrons and 
charged hadrons.

As already mentioned, a hadronic shower consists of a purely electromagnetic 
part and a purely hadronic one, with a relative ratio that depends on the 
energy of the incident hadron, and which is approximated by the function 
$f_{em}(E)$. The intrinsic difference in the response of a calorimeter to the 
two different parts of a hadronic shower is expressed by the $e/h$ ratio. 
Deviation of $e/h$ from 1 represents the difference in efficiency to
convert the deposited energy to visible signal, depending on the type of 
particles in the shower. This affects the energy resolution by a constant 
term, dependent on $e/h$,
\begin{eqnarray}
\left(\frac{\sigma(E)}{E}\right)_{e/h \neq 1} \propto |e/h - 1|
\end{eqnarray}
by a factor of the order of 15\%, 20\% 
\cite{ref:QC:pdg2000,ref:QC:wigmans91,ref:QC:wigmans87}.
In summary, all non statistical fluctuations contribute to the constant term 
of the resolution of a calorimeter. Construction defects, material 
inhomogeneity, calibration errors, non linear response of electronics etc., 
cause resolution degradation which is difficult to estimate in advance.

\section{Quartz fiber calorimetry}

\subsection{Motivation}

The basic motivation for developing new materials and techniques for 
detectors is the continuously increasing demands set by the experimental 
conditions at new accelerators. At the LHC, and also at the heavy ion physics
program at SPS and at RHIC, the experimental conditions at forward rapidity 
regions are very hostile for conventional materials that are used in 
calorimetry, mainly due to the very high radiation level. Besides, the 
calorimeters the signal of which is used in high level triggers must have 
fast response to cope with the high event rate in the experiments. For 
instance, at the LHC bunch crossing will be every 25~nsec for proton beams, 
125~nsec for $Pb$ beams. Calorimeters with quartz fibers fulfill both 
requirements of radiation resistance and fast response. First because quartz 
is a radiation hard material up to Grad level total dose. And second because 
the operation mechanism of quartz fiber calorimeters is based on the 
Cherenkov effect, which practically means that the generation and the 
duration of their signal do not last more than 10~nsec. In the following we 
discuss in detail the operation principle and the main properties of quartz 
fiber calorimeters
\cite{ref:QC:contin94}-\cite{ref:QC:hagopian99} and 
\cite{ref:QC:arnaldi98_3}-\cite{ref:QC:arnaldi98_2},
\cite{ref:QC:weber97,ref:QC:rhiczdc98,ref:QC:adler2000,ref:QC:andrieu2001}.

\subsection{Principle of operation and Cherenkov effect}
\label{subsec:QC:principle_of_operation}

The quartz fiber calorimeters are sampling calorimeters. The absorber is made 
of a dense material such as copper, iron, lead or tungsten and the sensitive
medium is composed of quartz fibers. The incident particles interact with the 
calorimeter's absorber and initiate showers. The charged particles of the 
shower traversing the optical fibers produce Cherenkov photons, which are 
guided along the fibers and are collected by photomultipliers. Production of 
Cherenkov radiation occurs when a charged particle travels in a medium with 
velocity higher than the velocity of light in that medium.
If $\beta=\upsilon/c$ is the velocity of the particle and $n$ is the index of 
refraction ($n$=$n(\lambda)$), then Cherenkov photons are emitted for
\begin{eqnarray}
\label{eq:QC:beta_thr}
\beta > \beta_{threshold} = \frac{1}{n}
\end{eqnarray}
at angle of emission $\theta_{c}$, relatively to the axis of motion of the 
particle. It holds
\begin{eqnarray}
\label{eq:QC:costheta}
\cos\theta_{c}=\frac{1}{n \beta }
\end{eqnarray}
The number of photons that are emitted per unit wavelength and per unit path 
length in the medium is given by the following equation
\begin{eqnarray}
\label{eq:QC:dN/dLdl}
\frac{d^{2}N_{ph}}{dL \ d\lambda}=2 \pi \alpha z^{2} \ \frac{\sin^{2}\theta_{c}}{\lambda ^{2}}
= 2 \pi \alpha z^{2} \ \frac{1}{\lambda^{2}} \cdot \left( 1-\frac{1}{n^2 \beta^2} \right)
\end{eqnarray}
where $\alpha$=1/137, $z$ is the particle charge in units of $e$, $\lambda$
the wavelength of the photon and $dL$ the path length in the medium. The 
emitted light is in the UV part of the spectrum due to the $1/\lambda^2$ 
production dependence. It should be noted that the yield of Cherenkov light 
is 2 to 3 orders of magnitude lower than the yield of scintillation light 
for the same amount of deposited energy. An intrinsic difference between the 
two effects is that scintillation light is emitted isotropically, whether 
Cherenkov light is emitted along a cone with opening angle $\theta_{c}$ with 
respect to the trajectory of the particle. In addition to that, the Cherenkov 
effect has a production threshold. 

Also, it should be emphasized that the Cherenkov effect occurs practically
simultaneously ($<$ 1~nsec) with the passage of particles and consequently 
the width of the produced signal is less than 10~nsec
\cite{ref:QC:gorodetzky95,ref:QC:britz95,ref:QC:akchurin97_399,ref:QC:arnaldi98_3}.

\begin{figure}[tb]
\centering
\begin{tabular}{cc}
\epsfxsize=190pt
\epsffile{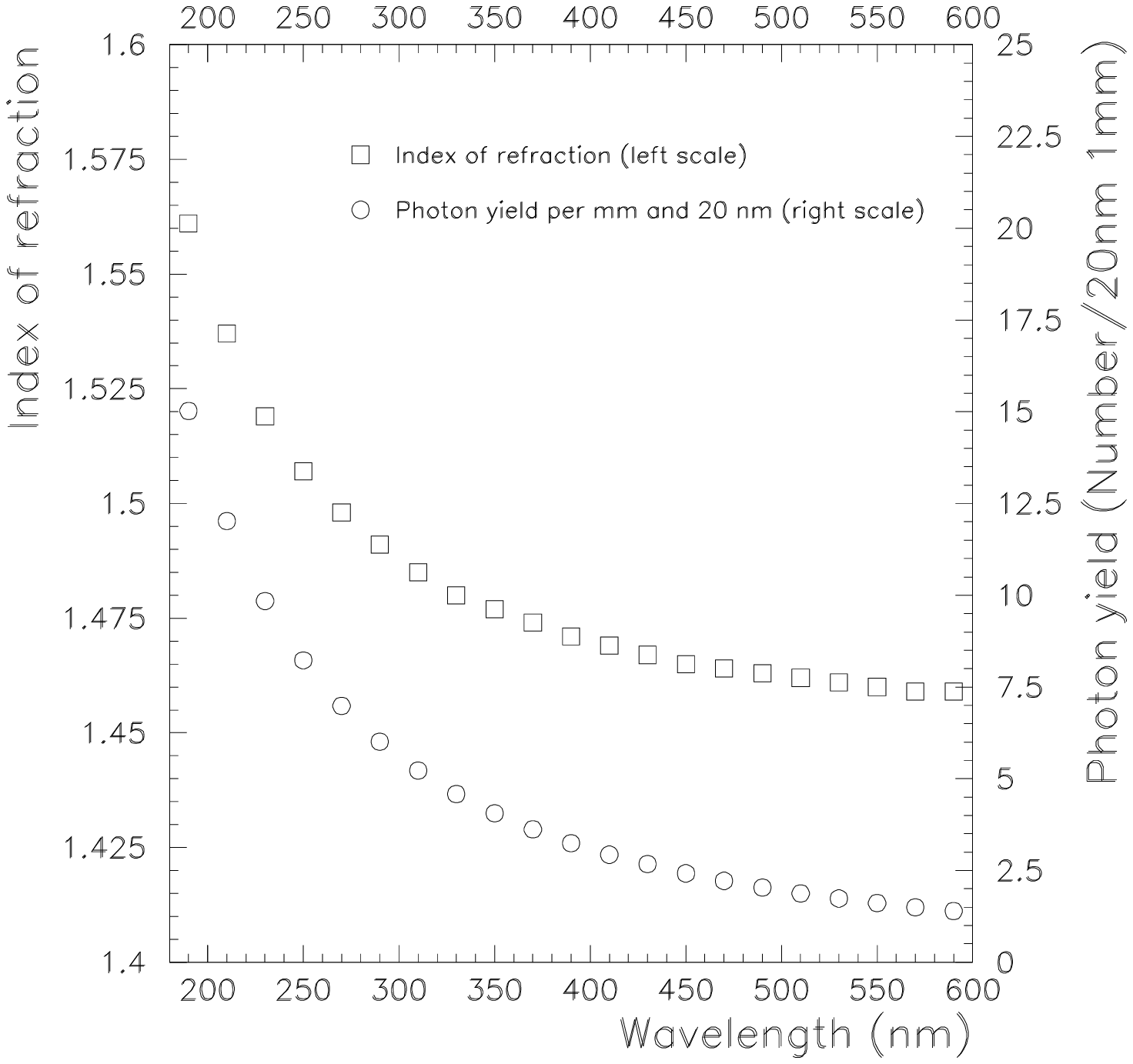}
\hspace*{10pt}&
\epsfxsize=190pt
\epsffile{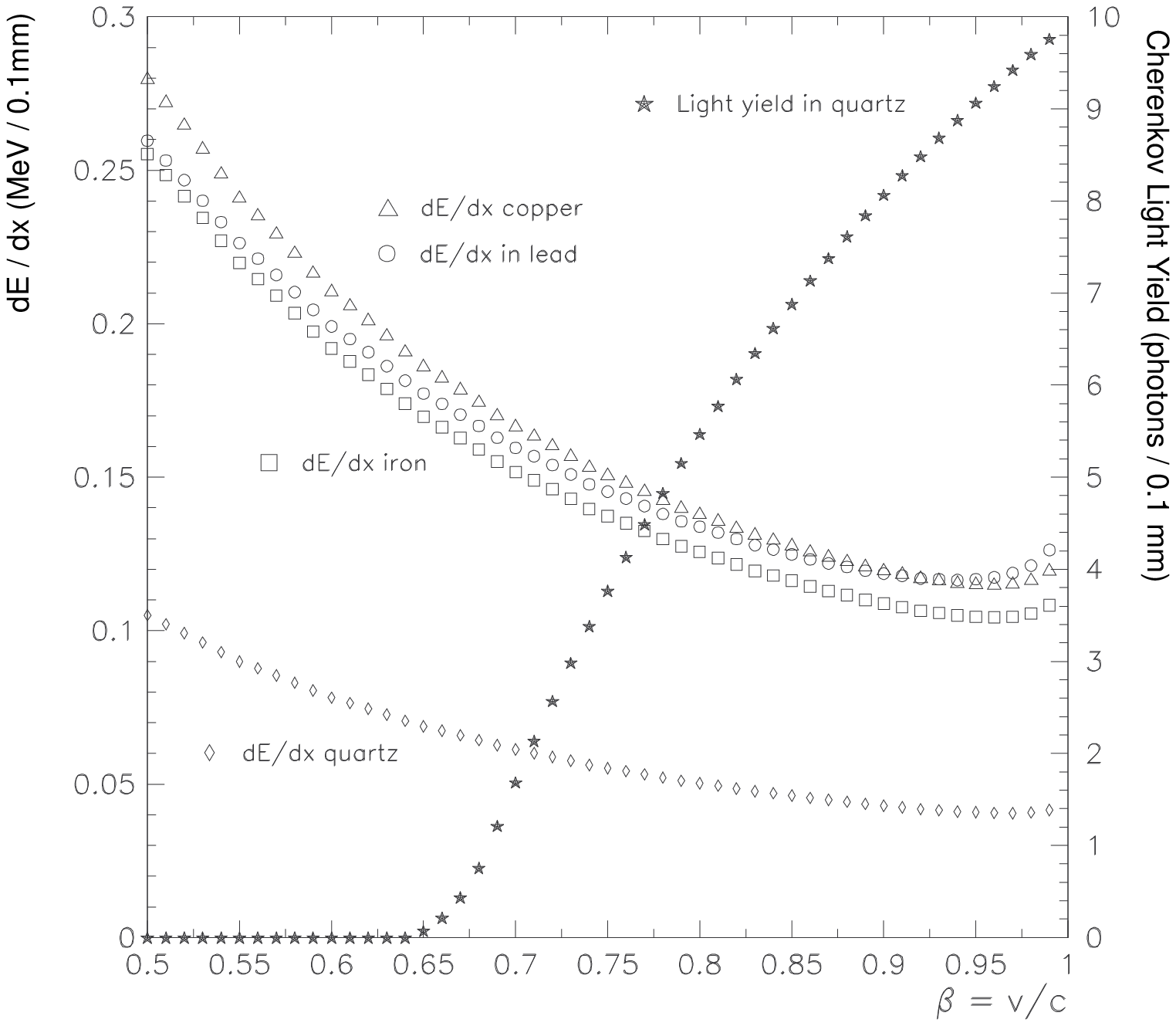}
\\
(a)  & (b)  \\
\end{tabular}
\caption{(a) index of refraction of quartz and Cherenkov photon production 
per unit length as a function of the wavelength, (b) energy loss per unit 
length from ionization in various materials and Cherenkov photon production 
per unit length in quartz as a function of the velocity of the incident 
particle.}
\label{fig:QC:refr_cher_de/dx}
\end{figure}

The index of refraction of quartz is in the range 1.46-1.55, for $\lambda$ 
600-200~nm, respectively (see fig.~\ref{fig:QC:refr_cher_de/dx}(a) from 
\cite{ref:QC:gorodetzky95} and also eq.~\ref{eq:QC:refr_index_quartz}).
Thus, the threshold velocity $\beta_{threshold}$ is 0.65-0.69 and the angle 
of emission $\theta_{c}$ is 46$^{\circ}$-50$^{\circ}$ for $\beta\simeq 1$. 
In table~\ref{tab:QC:refr_cher_particle}(a) index of refraction, velocity
and angle of emission for three different wavelengths are given. In 
table~\ref{tab:QC:refr_cher_particle}(b) we tabulate the energy required for 
$\beta=$ 0.65 and 0.99 for various charged particles. It results that the 
minimum kinetic energy that a charged particle should carry to produce 
Cherenkov light in quartz is 0.33~MeV for electrons and positrons, 119~MeV 
for pions and 802~MeV for protons.

\begin{table}[tb]
\centering
\caption{(a) index of refraction of quartz $n$, threshold velocity 
$\beta_{threshold}$ and angle of emission of Cherenkov photon 
$\theta_{c}$($\beta= 1$) for various wavelengths,
(b) total energy for $\beta$=0.65 and $\beta$=0.99 for various charged 
particles.}
\label{tab:QC:refr_cher_particle}
\vspace{10pt}
{\small
  \begin{tabular}{cccc}
  {\normalsize (a)}& \\
  \hline\hline
  $\lambda$(nm)    & $n$     & $\beta_{threshold}$ & $\theta_{c}$   \\
                   &         &                     & for $\beta= 1$ \\
  \hline
  & \\
               200 & 1.55051 & 0.645               & 49.8$^{\circ}$ \\
               400 & 1.46962 & 0.680               & 47.2$^{\circ}$ \\
               600 & 1.45840 & 0.686               & 46.7$^{\circ}$ \\
  & \\
  \hline\hline
  \end{tabular}
  }
  {\small
  \begin{tabular}{cccc}
  {\normalsize (b)}& \\
  \hline\hline
  particle      & mass     & E(MeV)             & E(MeV)           \\
                &(MeV)     & for $\beta$=0.65   & for $\beta$=0.99 \\
  \hline
  & \\
  $e^+,e^-$     &     0.5  &    0.6             &    3.5           \\
  $\pi^+,\pi^-$ &   139.6  &  183.7             &  989.6           \\
  $p$           &   938.3  & 1234.7             & 6651.4           \\
  & \\
  \hline\hline
  \end{tabular}
}
\end{table}

At this point we would like to remark a characteristic feature of quartz 
fiber calorimeters. From eq.~\ref{eq:QC:dN/dLdl}, we see that the particles 
of higher velocities produce more light and so their contribution to the 
total signal will be larger. In other words this means that a calorimeter 
which is based on quartz fibers is sensitive mainly to the core of the shower 
\cite{ref:QC:contin94,ref:QC:britz95,ref:QC:gorodetzky95,ref:QC:contin95,ref:QC:anzivino95p369}
since this part consists of charged particles with relatively larger energies 
and velocities. On the other hand the calorimeters that are based on 
$\frac{dE}{dx}$ technique are mainly sensitive to the charged particles of 
the shower with lower energies. Consequently the visible transverse size of 
electromagnetic and hadronic showers in quartz fiber calorimeters is narrower 
compared to the size of the showers visible by $\frac{dE}{dx}$ calorimeters.

The main difference between the two types of calorimeters is illustrated in 
fig.~\ref{fig:QC:refr_cher_de/dx}(b) (from \cite{ref:QC:gorodetzky95}) which 
depicts the Cherenkov light yield per unit length (eq.~\ref{eq:QC:dN/dLdl})
and the energy deposition per unit length ($dE/dx$ for various absorbers) as a 
function of particle velocity. It can be clearly seen that the significant part 
of the signal in conventional calorimeters is contributed by the lower energy 
particles of the shower since they deposit larger amount of energy.
Quartz fiber calorimeters do not collect the deposited energy on fibers
but the Cherenkov light emitted inside them by the relativistic charged 
particles of the shower.

\subsection{Light propagation in optical fibers}
\label{subsec:QC:lightguide_in_fibers}


An optical fiber consists of the core, with an index of refraction $n_{core}$, 
and the cladding, with index $n_{clad}$. The light propagates inside the 
fibers due to total internal reflection at the boundary surface of the two 
elements, and so it must be $n_{core} > n_{clad}$. When the light passes 
from a medium with index of refraction $n_{1}$ to a medium with $n_{2}$, it 
is reflected and refracted. It holds 
$\theta_{1}=\theta_{1^{'}},\ n_{1} \sin\theta_{1} = n_{2} \sin\theta_{2}$
with $\theta_{1}, \theta_{1^{'}}, \theta_{2}$ the angles between the normal 
to the boundary surface of the media and the directions of the incident, 
reflected and refracted light, respectively. The reflection is called 
internal or external if $n_{1} > n_{2}$ or $n_{1} < n_{2}$, respectively.
For $n_{1} > n_{2}$, and an incident angle $\theta \geq \arcsin(n_{2}/n_{1})$ 
we have reflection only, without refraction. This reflection is called 
{\em total internal reflection} and the minimum incident angle
required to occur is called {\em critical angle}. The critical angle for 
passing from a medium with $n_{1}$ to $n_{2}$ is 
$\theta_{critical}^{1\rightarrow 2} =\arcsin(n_{2}/n_{1})$. 
The so called {\em lightguide condition} in optical fibers is expressed by
\begin{eqnarray}
\label{eq:QC:transmission_condition}
\xi  \geq \arcsin\left(\frac{n_{clad}}{n_{core}}\right)
\end{eqnarray}
where $\xi$ is the angle of incidence of the light to the core-cladding 
boundary surface. Usually the optical fibers are characterized by the 
quantity NA, {\em numerical aperture}, which is defined as
\begin{eqnarray}
\label{eq:QC:na}
\textrm{NA} = \sqrt{n_{core}^2 - n_{clad}^2}
\end{eqnarray}
From eq.~\ref{eq:QC:transmission_condition} and eq.~\ref{eq:QC:na}
the lightguide condition can be written as follows
\begin{eqnarray}
\label{eq:QC:transmission_condition2}
\xi  \geq \arccos\left(\frac{\textrm{NA}}{n_{core}}\right)
& \textrm{or equivalently}
& \phi=90^{\circ} - \xi \leq \phi_{max}\equiv\arcsin\left(\frac{\textrm{NA}}{n_{core}}\right)
\end{eqnarray}
where $\phi$ is the angle between the direction of the light and the 
longitudinal axis of the fiber. For quartz fibers ($n_{core}\simeq$1.46), 
the angle $\phi_{max}$ is 8.7$^{\circ}$ for NA=0.22, 14.7$^{\circ}$ for 
NA=0.37.

\begin{figure}[tb]
\centering
\epsfysize=260pt
\epsffile{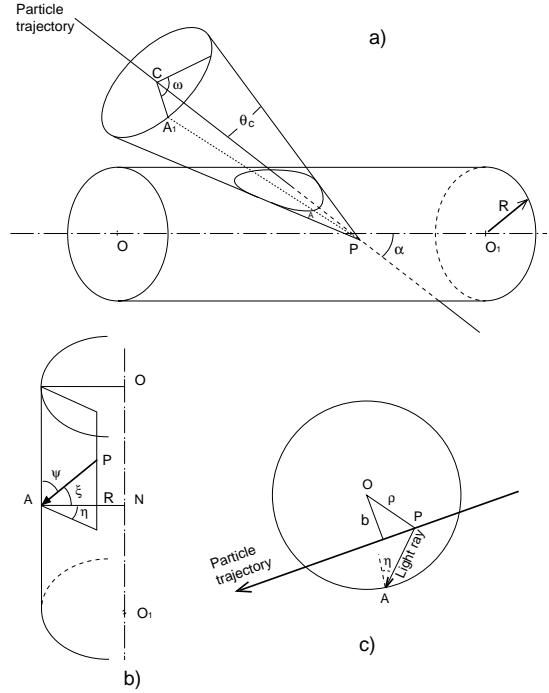}
\caption{schematic representation of Cherenkov photon production in an 
optical fiber and illustration of various geometrical parameters.}
\label{fig:QC:particle_cher_fiber}
\end{figure}

\begin{figure}[tb]
\centering
\epsfxsize=350pt
\epsffile{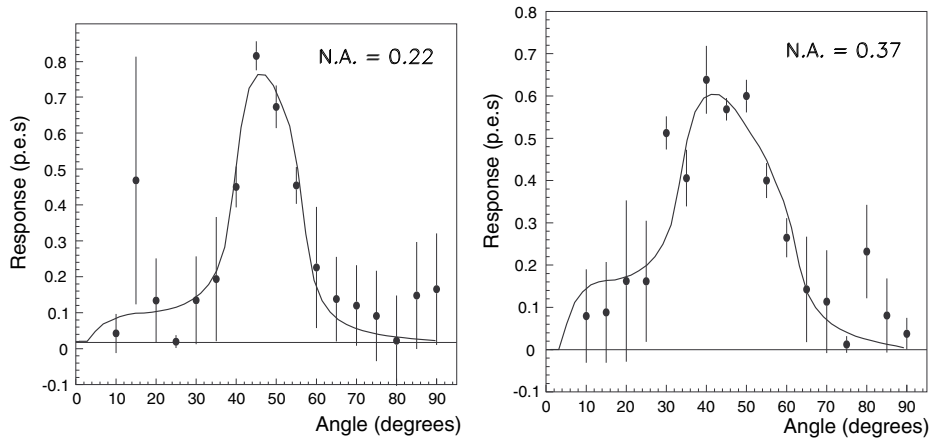}
\caption{experimental results on the dependence of Cherenkov light production 
on the angle of the incident particle with respect to fiber's central axis 
(fibers with NA=0.22 and NA=0.37).}
\label{fig:QC:yield_vs_angle}
\end{figure}

The Cherenkov light is emitted inside fibers at a specific angle with respect 
to the particle direction. The direction dependence of Cherenkov effect 
combined with the lightguide condition results to the fact that the light 
output depends on the angle at which the particles traverse the fibers.
In fig.~\ref{fig:QC:particle_cher_fiber} \cite{ref:QC:gorodetzky95} is shown 
a fiber of radius $R$ traversed by a particle whose track segment is at 
distance $b$ from the central axis and at angle $\alpha$. At point P, at 
distance $\rho$ from fiber's central axis, a Cherenkov photon is emitted and 
passes through the core-cladding boundary at angle $\xi$. We deduce:
\begin{eqnarray}
\label{eq:QC:cos_xi}
\cos\xi = \cos\eta \sin\psi
\end{eqnarray}
with
\begin{eqnarray}
\label{eq:QC:sin_eta}
 \sin\eta = \frac{\rho}{R} \sin\left(\arctan\frac{\sin\theta_{c} \sin\omega}{\cos\theta_{c} \sin\alpha
+ \cos\omega \sin\theta_{c} \cos\alpha} + \arcsin\frac{b}{\rho}\right)
\end{eqnarray}
and
\begin{eqnarray}
\label{eq:QC:cos_psi}
\cos\psi = \cos\alpha \cos\theta_{c} - \sin\alpha \sin\theta_{c} \cos\omega
\end{eqnarray}
From equations \ref{eq:QC:cos_xi}, \ref{eq:QC:sin_eta}, \ref{eq:QC:cos_psi} 
and the lightguide condition of eq.~\ref{eq:QC:transmission_condition2} we 
have that the production\footnote[1]{With production we mean initial 
production, capture and successful transmission of photons to the edge of 
fibers in order to be detected and collected as signal.} of Cherenkov light 
is maximum for particles traversing the quartz fibers at angles of incidence 
in the range $\alpha \pm \Delta\alpha$ with 
$\alpha \simeq \theta_{c}\sim$ 40$^{\circ}$ 
to 50$^{\circ}$ and $\Delta\alpha \simeq \phi_{max}$. 
In fig.~\ref{fig:QC:yield_vs_angle} \cite{ref:QC:gorodetzky95} the 
experimental confirmation of this is illustrated
\cite{ref:QC:contin94,ref:QC:britz95,ref:QC:lundin96,ref:QC:anzivino95p237,ref:QC:anzivino95p369,ref:QC:adler2000}.

\subsection{Quartz optical fibers}

The fibers of this type are composed of synthetic fused silica,
term commonly used for SiO$_2$ in non crystallic state (amorphous material).
It is formed by chemical combination of silicon with oxygen in special 
production process, which provides high purity, level of impurities of the 
order of 1 part per million. It should not be confused with the fused silica or 
fused quartz which is a material produced by compression and fusion of 
crystals of mineral silica. This, besides SiO$_2$, contains oxides of 
various elements such as Fe$_2$O, Al$_2$O$_3$, TiO$_2$, CaO, MgO, Na$_2$O. 
For clarification it should be stressed that by the terms quartz or pure 
quartz, used often in literature of calorimeters, we mean synthetic fused 
silica.

Synthetic fused silica has fine optical properties and is widely used in 
industry of optical systems for producing high quality lenses. It is 
transparent in a wide range of spectrum, from UV region to IR, has low 
thermal expansion coefficient ($\sim 5.5 \cdot 10^{-7}/^{\circ}$C) and very 
low coefficient of refractive index change with temperature 
($\sim 1.3 \cdot 10^{-5}/^{\circ}$C). Its dispersion relation is 
approximated for 20$^{\circ}$C by the following equation 
\cite{ref:QC:malitson65}, with precision in $n$ of $\pm 3 \cdot 10^{-5}$,
\begin{eqnarray}
\label{eq:QC:refr_index_quartz}
n^2 -1 = \frac{0.6961663 \lambda^2}{\lambda^2 - (0.0684043)^2} + \frac{0.4079426 \lambda^2}{\lambda^2 - (0.1162414)^2}
+ \frac{0.8974794 \lambda^2}{\lambda^2 - (9.896161)^2} \ ,   \lambda \ \textrm{in} \  \mu\textrm{m}
\end{eqnarray}
The quartz optical fibers are made of core composed of high purity synthetic 
fused silica or doped with hydroxyl (OH$^-$) in high ($\sim$ 1000 ppm) or 
low content ($\sim$ 10 ppm). Their cladding can be composed of quartz doped 
with fluorine, or is a polymer (hard plastic). Quartz fibers are available 
in core diameters of 200 to 1000~$\mu$m, cladding width about 50~$\mu$m and 
numerical aperture NA=0.22, 0.37 or about 0.50. Light of wavelength from 
$\sim$~200~nm to $\sim$~1200~nm is transmitted inside them with attenuation 
less than 0.05~dB/m, corresponding to an attenuation length 
$l_{att} \sim$~90~m. This type of fibers is used in medical and industrial 
applications, mainly for light transmission or collection in IR.

At this point, it should be also mentioned that quartz, in addition to above 
properties, is also characterised by extreme resistance to high radiation 
dose (of the order of 10~Grad) \cite{ref:QC:gorodetzky93}, a feature which 
derives from the strong Si-O chemical bond.
Thus the use of fibers with core and cladding composed of quartz 
(quartz core + fluorine doped quartz cladding)
\cite{ref:QC:fabian90,ref:QC:gavrilov94,ref:QC:avezov97,ref:QC:hagopian99}
is the only solution for calorimeters that operate in high radiation 
environment, and in general very hostile conditions for conventional 
materials. A disadvantage of this type of fibers is their relative high cost. 
In addition it is a material of high hardness, a feature which makes 
necessary the use of special shearing methods.

\subsection{Basic properties of quartz fiber calorimeters}
\label{subsec:QC:basic_properties}

\begin{figure}[tbp]
\centering
\epsfysize=220pt
\epsffile{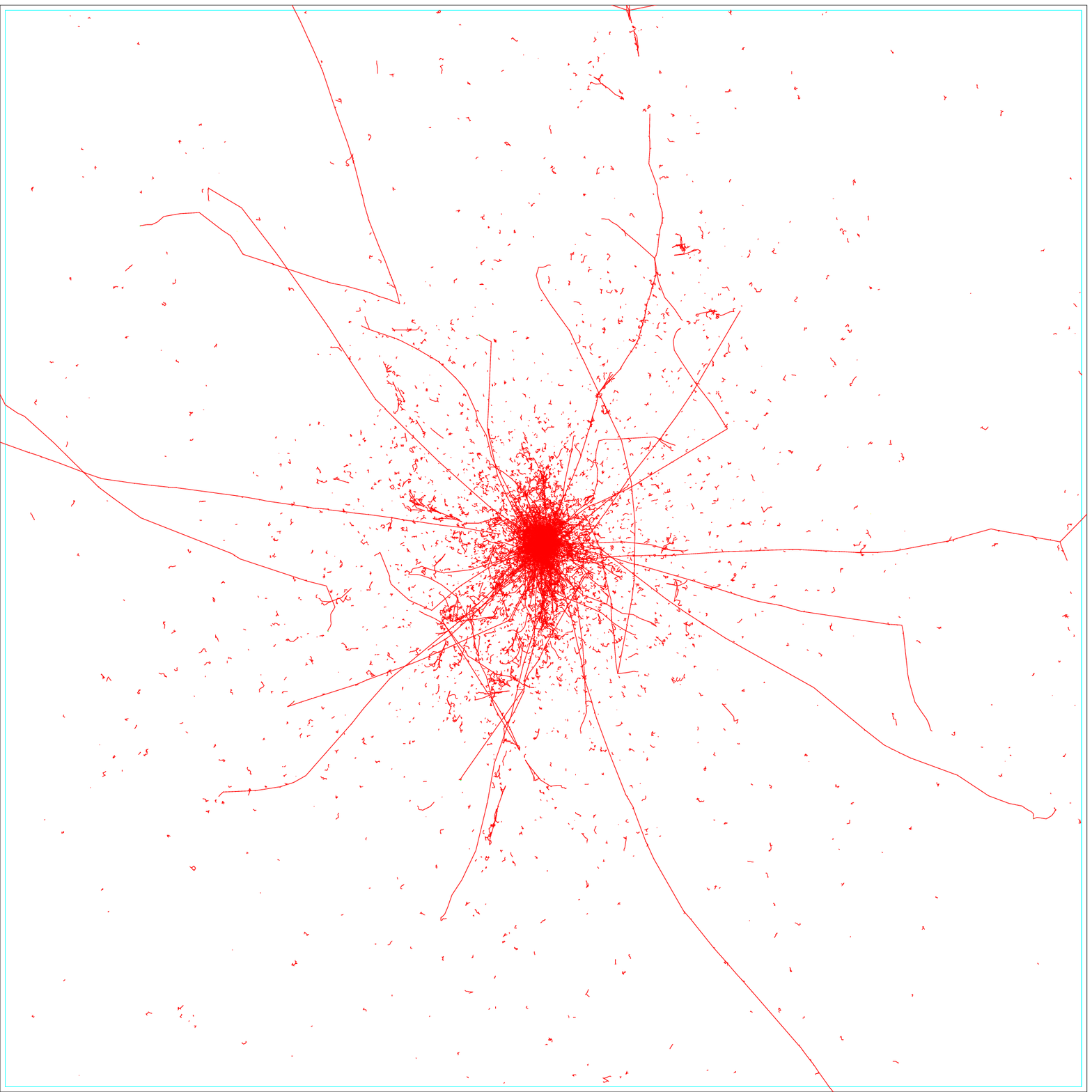}
\caption{front projected view of a 100 GeV charged pion shower where all 
charged particle tracks are shown (square size is 20$\times$20 cm$^2$, 
absorber is W). This is the part of the shower that a $\frac{dE}{dx}$ based 
calorimeter is sensitive to.}
\label{fig:QC:width_dedx}

\vspace{30pt}

\epsfysize=220pt
\epsffile{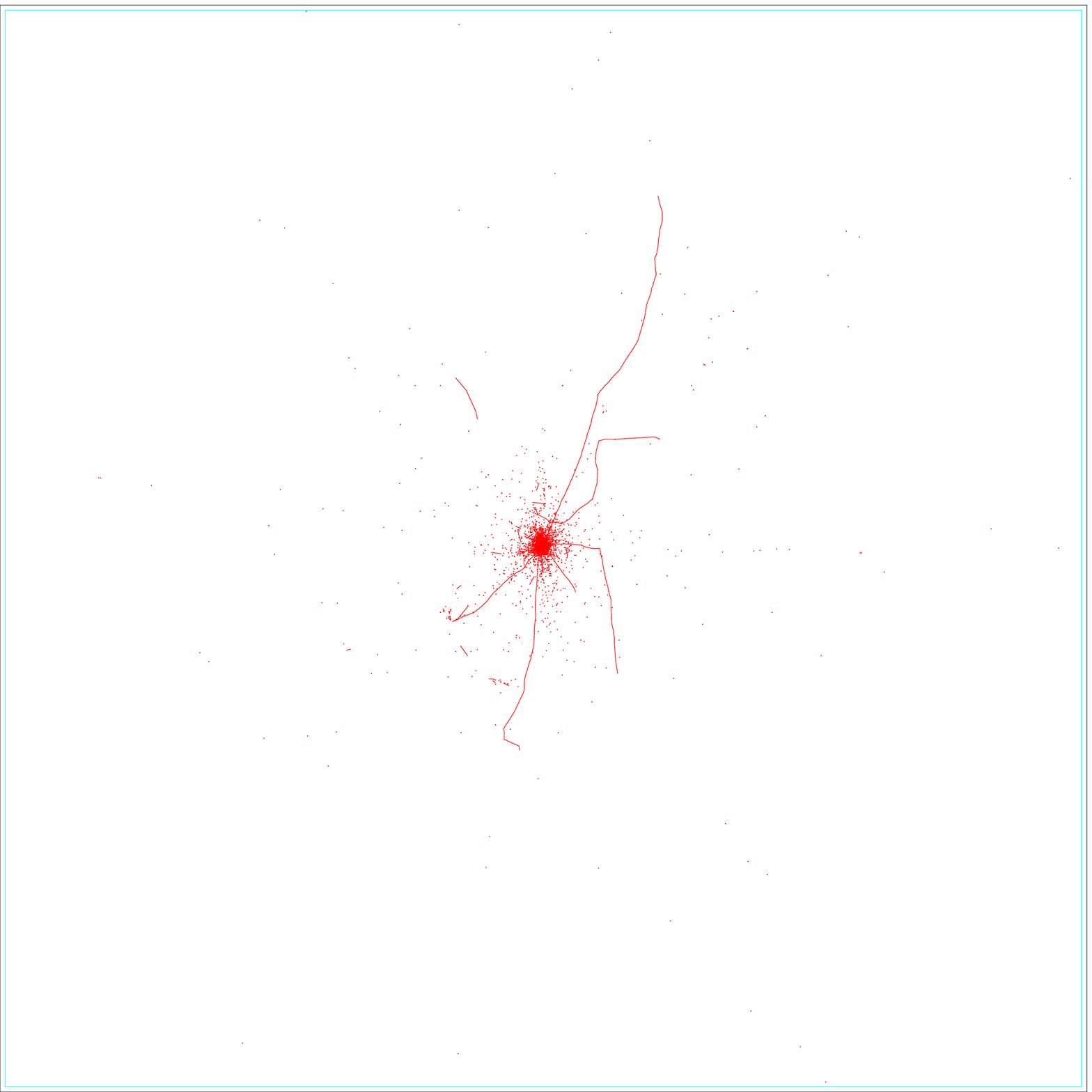}
\caption{front projected view of a 100 GeV charged pion shower where only 
tracks of charged particles with $\beta\geq\beta_{threshold}$ and with 
direction that gives high probability of Cherenkov photon capture in fibers 
are shown (square size is 20$\times$20 cm$^2$, absorber is W, fibers with 
numerical aperture NA=0.40 and 45$^\circ$ inclination). This is the part of 
the shower that a quartz fiber calorimeter is sensitive to.}
\label{fig:QC:width_qcal}
\end{figure}

We summarize the basic properties and the main advantages of quartz fiber 
calorimetry that have been mentioned in previous subsections. The signal in 
this type of calorimeters is the Cherenkov light which is produced inside the 
fibers by the traversing charged particles of the shower with 
$\beta > \beta_{threshold}\simeq 0.7$. The fact that for higher particle 
velocity more Cherenkov photons are emitted, in combination with the 
directionality of the effect, due to specific angle of  emission, and the 
limitations imposed by the lightguide condition in fibers, make quartz fiber 
calorimeters be mainly sensitive to the central part of the shower. In other 
words, the visible transverse width of electromagnetic and hadronic showers 
is very narrow, in transverse width of about 10~mm and 5-10~cm, respectively,
95\% of the signal is contained
\cite{ref:QC:contin94,ref:QC:britz95,ref:QC:gorodetzky95,ref:QC:contin95,ref:QC:anzivino95p369} and
\cite{ref:QC:chiavassa95,ref:QC:akchurin97_399,ref:QC:arnaldi2001,ref:QC:weber97,ref:QC:andrieu2001}.
This size is 3 to 4 times narrower compared to that of calorimeters operating with 
the $\frac{dE}{dx}$ technique. This is well illustrated in 
fig.~\ref{fig:QC:width_dedx} and~\ref{fig:QC:width_qcal} 
\cite{ref:QC:gm2,ref:QC:gm_phd}, which show the part of hadronic shower that 
a $\frac{dE}{dx}$ and a quartz fiber based calorimeter is sensitive to, 
respectively. As a rule of thumb, one can state that the visible part of 
hadronic showers in quartz fiber calorimeters, and especially for those with 
45$^\circ$ fiber inclination geometry, has a transverse development which 
seems to scale with the Moliere radius and not with the interaction length, 
as in conventional $\frac{dE}{dx}$ calorimeters.
On the other hand, since the signal of the hadronic showers is produced 
mainly by their electromagnetic component, the quartz fiber calorimeters 
are non-compensating ($e/h > 1$).

Another characteristic feature of quartz fiber calorimeters is their very 
fast response. The Cherenkov effect is an intrinsically fast process which 
occurs almost simultaneously with the passage of particles. Consequently 
the duration of signal generation and the signal width do not exceed 10~nsec 
in time \cite{ref:QC:britz95,ref:QC:gorodetzky95,ref:QC:arnaldi98_3,ref:QC:akchurin97_399}.

Using fibers with quartz core and quartz cladding, we can build very radiation 
hard calorimeters capable to operate up to total dose level of the order of 
10~Grad \cite{ref:QC:gorodetzky93,ref:QC:gavrilov94,ref:QC:avezov97,ref:QC:hagopian99}.

The main disadvantage of this calorimetric technique is the low light yield, 
2 to 3 orders of magnitude lower compared to scintillator based calorimeters 
for the same amount of deposited energy. To maximize the light production, 
a geometry with fiber inclination at $45^{\circ}$ with respect to the 
direction of the incident particles is required. With such a configuration a 
better exploitation of the directionality of the Cherenkov effect and the 
lightguide condition in fibers is achieved
\cite{ref:QC:contin94,ref:QC:britz95,ref:QC:lundin96,ref:QC:anzivino95p237,ref:QC:anzivino95p369,ref:QC:adler2000}.
A $45^{\circ}$ inclination geometry sets restrictions and difficulties in 
construction and is a considerable limitation factor, if we require the 
calorimeter to provide position measurement as well. Calorimeters with 
spaghetti structure, where the fibers are embedded inside the absorber volume 
with a $0^{\circ}$ inclination angle with respect to incoming particles, have 
satisfying light yield for incident particles or jets with energy at the 
TeV scale. Also, it should be mentioned that quartz fibers are about an order of 
magnitude more expensive than scintillating fibers, so in general it is not 
affordable to increase the light production by increasing the filling ratio 
of the calorimeter.

We conclude this section with table~\ref{tab:QC:qcal_dE/dx}, where the main 
features and properties of quartz fiber calorimetry in comparison with those 
for $\frac{dE}{dx}$ method are summarized 
\cite{ref:QC:contin94}-\cite{ref:QC:hagopian99}.

\begin{table}[tbp]
\centering
\caption{main features of quartz fiber calorimetry in comparison with
 $\frac{dE}{dx}$ calorimetry.}
\label{tab:QC:qcal_dE/dx}
\vspace{10pt}
{\footnotesize
\begin{tabular}{ll}
\hline\hline 
                                                        & \\
{\small\bf quartz fiber calorimetry}                    & {\small\bf $\frac{dE}{dx}$ calorimetry} \\
                                                        & \\
\hline 
                                                        & \\
signal is the Cherenkov light produced                  & signal (light or charge) produced by shower  \\
in fibers by the charged particles of the               & energy deposited in active material  \\
shower                                                  & \\ 
                                                        & \\
Cherenkov effect has a threshold ($\beta>\frac{1}{n}$)  & no threshold \\
                                                        & \\
non isotropic light emission                            & light emitted isotropically\\
                                                        & \\
sensitive mainly to relativistic charged                & sensitive mainly to low energy charged  \\
particles of shower                                     & particles of shower                 \\
                                                        & \\
intrinsically insensitive to low energy                 & common source of background\\
neutrons and to radioactivity                           & \\
                                                        & \\
detector sensitive to shower core                       & detector sensitive to total shower\\
                                                        & \\
visible transverse size of hadronic shower              & \\
for 95\% signal containement $\cal O$(5-10~cm),         & 3 to 4 times wider  \\
$\cal O$(10~mm) for electromagnetic shower              & \\
                                                        & \\
light yield $\cal O$(10~photoelectrons/GeV)             & $\cal O$(1000~photoelectrons/GeV) for a  \\
for $45^{\circ}$ inclination geometry,                  & scintillation based calorimeter \\
$\cal O$(1~p.e./GeV) for $0^{\circ}$ geometry           & \\
                                                        & \\
quartz fibers retain their optical properties           & sensitive materials with considerable degradation  \\
after total radiation dose of the order of              & of their properties or total destruction to    \\
1 Grad                                                  & radiation level of 0.01-0.1~Grad\\
                                                        & \\
calorimeter capable to operate in very forward          & detector to operate in less demanding  \\
rapidity region, in high radiation environment,         & conditions \\
in high event rate experiments                          & \\
                                                        & \\
\hline\hline 
\end{tabular}
}
\end{table}

\section{Quartz fiber calorimeters}

The main advantages of quartz fiber calorimeters are, the radiation hardness, 
to very high dose level (of the order of 1~Grad or more), their fast response 
($< 10$~nsec) and the compact detector dimensions, since the transverse size 
of the visible shower is very narrow. Calorimeters of this type operate as 
centrality detectors, trigger detectors, luminosity monitors or calorimeters 
at very forward regions. In the following subsections we describe the quartz 
fiber calorimeters that are in operation or under construction/development in 
various experiments. We first present the calorimeters with fibers placed at 
$0^{\circ}$ inclination with respect to beam axis (``spaghetti'' geometry), 
and then we continue with the ones with $45^{\circ}$ construction geometry 
(``sandwich'' structure). The specifications of each calorimeter are tabulated 
in tables \ref{tab:QC:qcal_na50} to \ref{tab:QC:castor}, their main parameters 
are summarized in tab.~\ref{tab:QC:all_qcals}.

\subsection{The Zero Degree calorimeter for the NA50 experiment at CERN-SPS}

\begin{figure}[f]
\centering
\rotatebox{-90}{%
\epsfysize=250pt
\epsffile{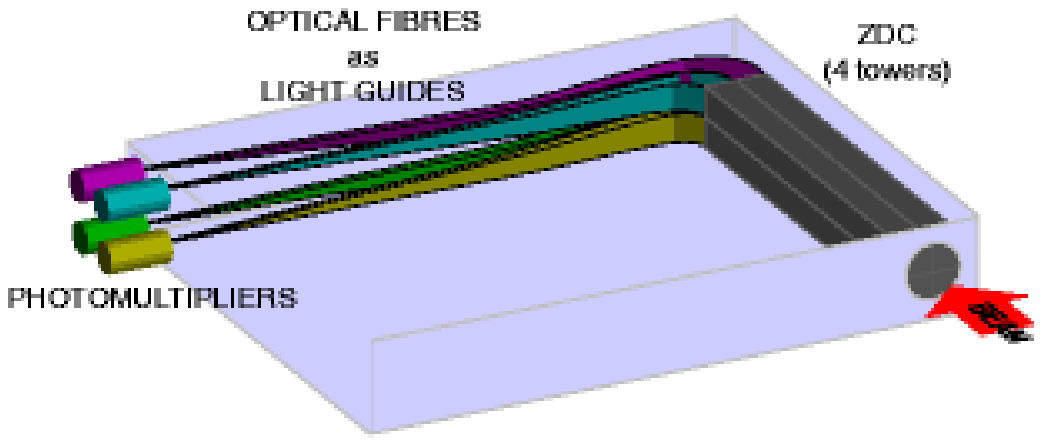}
}%
\caption{general view of the Zero Degree calorimeter for the NA50 experiment.}
\label{fig:QC:qcal_na50}
\end{figure}

The NA50 \cite{ref:QC:na50,ref:QC:na50_2000} is a fixed target experiment
of the heavy ion physics research program at CERN-SPS, for detection of matter 
in quark-gluon plasma phase. The purpose of the experiment is to study the 
production of vector mesons, $\rho, \omega, \phi, J/\Psi$ in heavy ion 
collisions (heavy ion beam on lead target) up to 158~GeV/nucleon beam energy.
A quartz fiber calorimeter is used to determine the centrality of the 
collisions. The calorimeter is placed at 1.6~m after the target, covering 
pseudorapidity range $\eta \geq 6.3$ (Zero Degree calorimeter)
\cite{ref:QC:arnaldi98_3}-\cite{ref:QC:anzivino94}, 
\cite{ref:QC:arnaldi98_1,ref:QC:arnaldi98_2}.
A calorimeter with quartz fibers is imposed by the requirements of tolerance 
to very high radiation dose expected on the detector site, of the order of 
10~Grad, and of fast response, since it provides the trigger signal for the 
experiment. The calorimeter has a spaghetti structure with fibers embedded 
in absorber volume at $0^{\circ}$ inclination with respect to the beam axis. 
A brief description of its elements and its main properties is given in 
table~\ref{tab:QC:qcal_na50}. In fig.~\ref{fig:QC:qcal_na50} a general view 
of the NA50 ZDC is shown.

\begin{table}[f]
\centering
\caption{specifications of the Zero Degree calorimeter for the NA50 experiment.}
\label{tab:QC:qcal_na50}
\vspace{10pt}
{\footnotesize
\begin{tabular}{ll}
\hline\hline
                   & \hfill \ \\
{\bf  purpose}     & measurement of energy of spectator nucleons (and nuclear fragments) to \\
                   & determine impact parameter of collision and number of participant nucleons\\
                   & \\
{\bf  position}    & 1.6~m after the target, covering pseudorapidity range $\eta \geq 6.3$ \\
                   & \\
{\bf  construction}& $5\times5\times65$~cm$^3$ segmented in 4 towers, depth 5.6~$\lambda_{I}$,\\
                   & fibers at $0^{\circ}$ inclination embedded in grooved absorber plates, \\
                   & 1 photomultiplier per tower (total: 4)\\
                   & \\
{\bf  absorber}    & tantalum (Ta: $\lambda_{I}$= 11.5~cm, $X_0$= 0.4 cm, density= 16.65~gr/cm$^3$),\\
                   & 30 plates, thickness 1.5~mm each  \\
                   & \\
{\bf  fibers}      & pure silica core ($\varnothing$ 0.365~mm), silica fluorinated cladding,\\
                   & numerical aperture NA=0.22 \\
                   & 900 fibers uniformly distributed with 1.5 mm pitch \\
                   & \\
{\bf filling ratio}&$\frac{\textrm{fiber volume}}{\textrm{absorber volume}}$ : 1/17 (=5.88\%) \\
                   & \\
{\bf $\sigma$(E)/E}& 35(28)\% at 100(205) GeV (= $\frac{2.9}{\sqrt{E(GeV)}} \oplus 19 \%$) \\
                   & \\
{\bf light yield}  & 0.5 photoelectrons per GeV (protons)\\
                   & \\
{\bf total dose}   & 10 Grad \\
                   & \\
\hline\hline
\end{tabular}
}
\end{table}

\subsection{The Very Forward calorimeter for the CMS experiment at CERN-LHC}
\label{subsec:QC:cms_qcal}

\begin{figure}[p]
\centering
\rotatebox{-90}{%
\epsfxsize=250pt
\epsffile{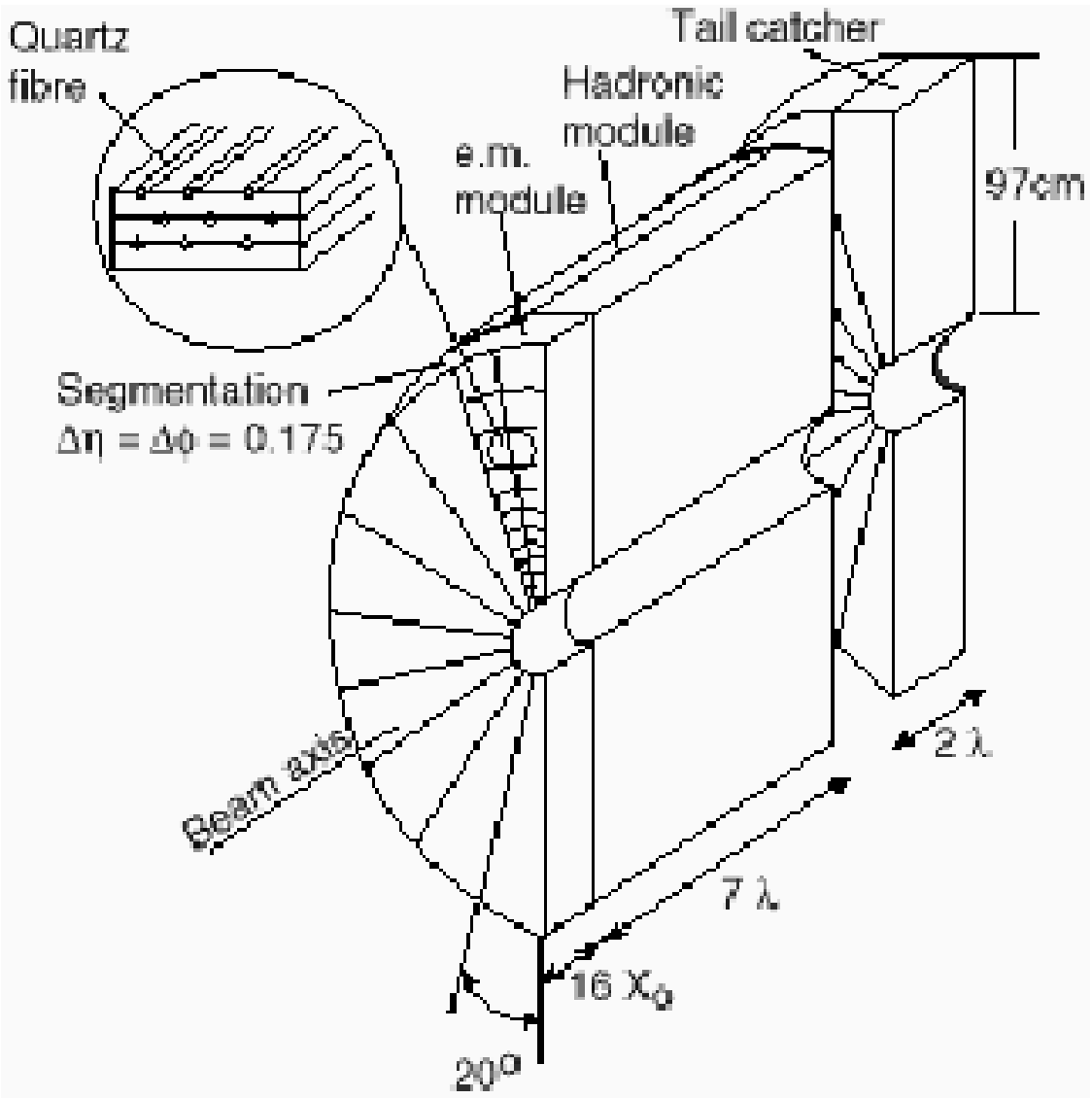}
}%
\caption{schematic view of one half of the CMS Very Forward calorimeter.}
\label{fig:QC:qcal_cms}
\end{figure}

The CMS \cite{ref:QC:cms} is a collider experiment at the LHC. It will
perform Higgs boson and supersymmetric particles research in $p p$
collisions at $\sqrt{S}_{p+p} = 14$~TeV. In the very forward region 
(3 $\leq |\eta| \leq$ 5) a calorimeter \cite{ref:QC:onel2002}-\cite{ref:QC:ferrando97}
with quartz fibers will be used in order to cope with the high 
radiation and the high event rate requirements of operation. The calorimeter 
has a spaghetti structure, with fibers embedded in absorber volume at 
$0^{\circ}$ inclination with respect to beam axis. Two identical calorimeters
are being built, placed symmetrically in forward and backward regions, and at 
distance of 11~m from the interaction point. They provide $2\pi$ azimuth 
coverage and are segmented in towers with $\Delta\eta=\Delta\phi=0.175$.
The specifications of the CMS Very Forward calorimeter are tabulated in
table~\ref{tab:QC:qcal_cms}, a schematic view is shown in 
fig.~\ref{fig:QC:qcal_cms}.

\begin{table}[p]
\centering
\caption{specifications of the Very Forward calorimeter for the CMS experiment.}
\label{tab:QC:qcal_cms}
\vspace{10pt}
{\footnotesize
\begin{tabular}{ll}
\hline\hline
                   & \hfill \ \\
{\bf  purpose}     & missing energy measurement and forward jet tagging\\
                   & \\
{\bf  position}    & $\pm$11~m from the interaction point, covering 3 $\leq |\eta| \leq$ 5 in pseudorapidity\\
                   & and $\phi=2\pi$ in azimuth \\
                   & \\
{\bf  construction}& segmented in towers with $\Delta \eta = \Delta \phi = 0.175$, depth 8.8~$\lambda_{I}$,\\
                   & fibers at $0^{\circ}$ inclination embedded in grooved absorber plates \\
                   & 1 photomultiplier per tower (total: 360)\\
                   & \\
{\bf  absorber}    & steel ($\lambda_{I}$= 17.4~cm, $X_0$= 1.9~cm, density= 7.5~gr/cm$^3$)\\
                   & \\
{\bf  fibers}      & quartz core ($\varnothing$ 0.300~mm), fluorinated doped quartz cladding,\\
                   & numerical aperture NA=0.22 \\
                   & \\
{\bf filling ratio}&$\frac{\textrm{fiber volume}}{\textrm{absorber volume}}$ $<$ 1\% \\
                   & \\
{\bf $\sigma$(E)/E}& $\frac{2.7}{\sqrt{E(GeV)}} \oplus 13\%$ \\ 
                   & \\
{\bf light yield}  & $\sim$ 0.5 photoelectrons per GeV \\ 
                   & \\
{\bf total dose}   & 1 Grad \\
                   & \\
\hline\hline
\end{tabular}
}
\end{table}

\subsection{The Zero Degree calorimeters for the ALICE experiment at CERN-LHC}
\label{subsec:QC:alice_zdc}

\begin{figure}[p]
\centering
\rotatebox{-90}{%
\epsfxsize=250pt
\epsffile{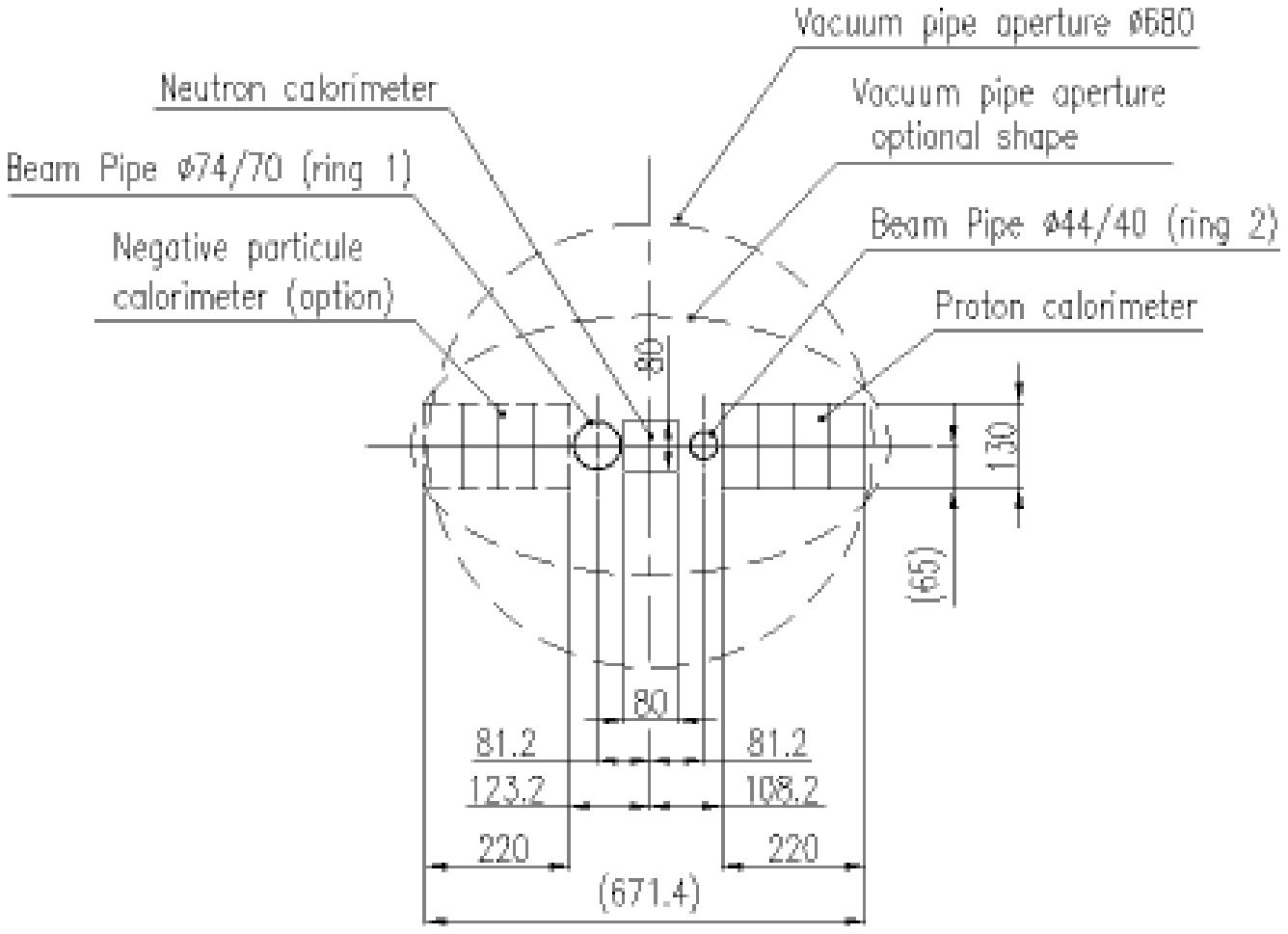}
}%
\caption{cross section of the LHC beam line and the Zero Degree calorimeters of the
ALICE experiment.}
\label{fig:QC:qcal_alicezdc}
\end{figure}

The ALICE~\cite{ref:QC:alice_proposal} is a heavy ion physics dedicated 
collider experiment at the LHC. It is designed to search simultaneously for 
all observables that have been proposed as characteristic signatures of 
production of matter in quark-gluon plasma state. It will investigate various 
systems produced in heavy ion collisions up to a center of mass energy of
$\sqrt{S}_{Pb+Pb} = 1148$~TeV. Zero degree calorimeters that measure the 
energy of incident neutrons and protons which do not participate in the 
collision (spectator nucleons) will be used to determine the centrality of 
the heavy ion collisions. The experimental conditions impose the use of a 
detector that comprises very fast response, since its signal serves as a 
trigger signal for the other detectors of the experiment, compact physical 
dimensions, and high resistance to radiation. These requirements make a 
quartz fiber calorimeter be the only solution. The ALICE will use two pairs 
of neutron and proton quartz fiber calorimeters, positioned symmetrically on 
both sides of the interaction point and at distance of about 115~m 
\cite{ref:QC:alicezdc_tdr}-\cite{ref:QC:arnaldi98_2}.
Both calorimeters have spaghetti geometries, with fibers embedded in absorber
at $0^{\circ}$ inclination with respect to the beam axis. A brief description
of their construction and properties is given in 
table~\ref{tab:QC:qcal_alicezdc}. In fig.~\ref{fig:QC:qcal_alicezdc} a view 
of the position of the calorimeters between the beam pipes of the LHC is shown.

\begin{table}[p]
\centering
\caption{specifications of the Zero Degree calorimeters for the ALICE experiment.}
\label{tab:QC:qcal_alicezdc}
\vspace{10pt}
{\scriptsize
\begin{tabular}{lll}
\hline\hline
                   & & \\
                   & {\bf\small neutron ZDC }                                 & {\bf\small proton ZDC }\\
                   & & \\
{\bf  purpose}     & measurement of spectator neutrons energy                 & measurement of spectator protons energy\\
                   & for impact parameter determination                       & for impact parameter determination  \\
                   & & \\
{\bf  position}    & $\pm$116.13~m from the interaction point                 & $\pm$115.63~m from the interaction point\\
                   & & \\
{\bf  construction}& $7\times7\times100$~cm$^3$, depth 8.5~$\lambda_{I}$,     & $20.8\times12\times150$~cm$^3$, depth 8.4~$\lambda_{I}$,\\
                   & fibers at $0^{\circ}$ inclination,                       & fibers at $0^{\circ}$ inclination,\\
                   & 5 photomultipliers in total                              & 5 photomultipliers in total \\
                   & & \\
{\bf  absorber}    & W alloy or tantalum (Ta: $\lambda_{I}$= 11.5~cm,         & brass ($\lambda_{I}$= 18.4~cm, $\rho$= 8.48~gr/cm$^3$),\\
                   & $X_0$= 0.4~cm, $\rho$= 16.65~gr/cm$^3$),                 & 52 plates 4~mm thick each\\
                   & 44 plates 1.6~mm thick each                              & \\
                   & & \\
{\bf  fibers}      & pure silica core ($\varnothing$ 0.365~mm),               & pure silica core ($\varnothing$ 0.550~mm),\\
                   & silica fluorinated cladding,                             & silica fluorinated cladding, \\
                   & numerical aperture NA=0.22                               & numerical aperture NA=0.22 \\
                   & & \\
{\bf filling ratio}&$\frac{\textrm{fiber volume}}{\textrm{absorber volume}}$ : 1/22 (=4.55\%) & 1/65 (=1.54\%) \\
                   & & \\
{\bf $\sigma$(E)/E}& 10.5\% for 2.7~TeV neutrons                              & $\sim$10\% for 2.7~TeV protons \\
                   & & \\
{\bf light yield}  & 0.35 photoelectrons per GeV                              & 0.28 photoelectrons per GeV\\
                   & \\
{\bf total dose}   & $\cal O$(1) Grad                                         & $\cal O$(1) Grad \\
                   & & \\
\hline\hline
\end{tabular}
}
\end{table}

\subsection{The Very Forward EM calorimeter for the NA52 experiment at CERN-SPS}

\begin{figure}[p]
\centering
\rotatebox{-90}{%
\epsfxsize=250pt
\epsffile{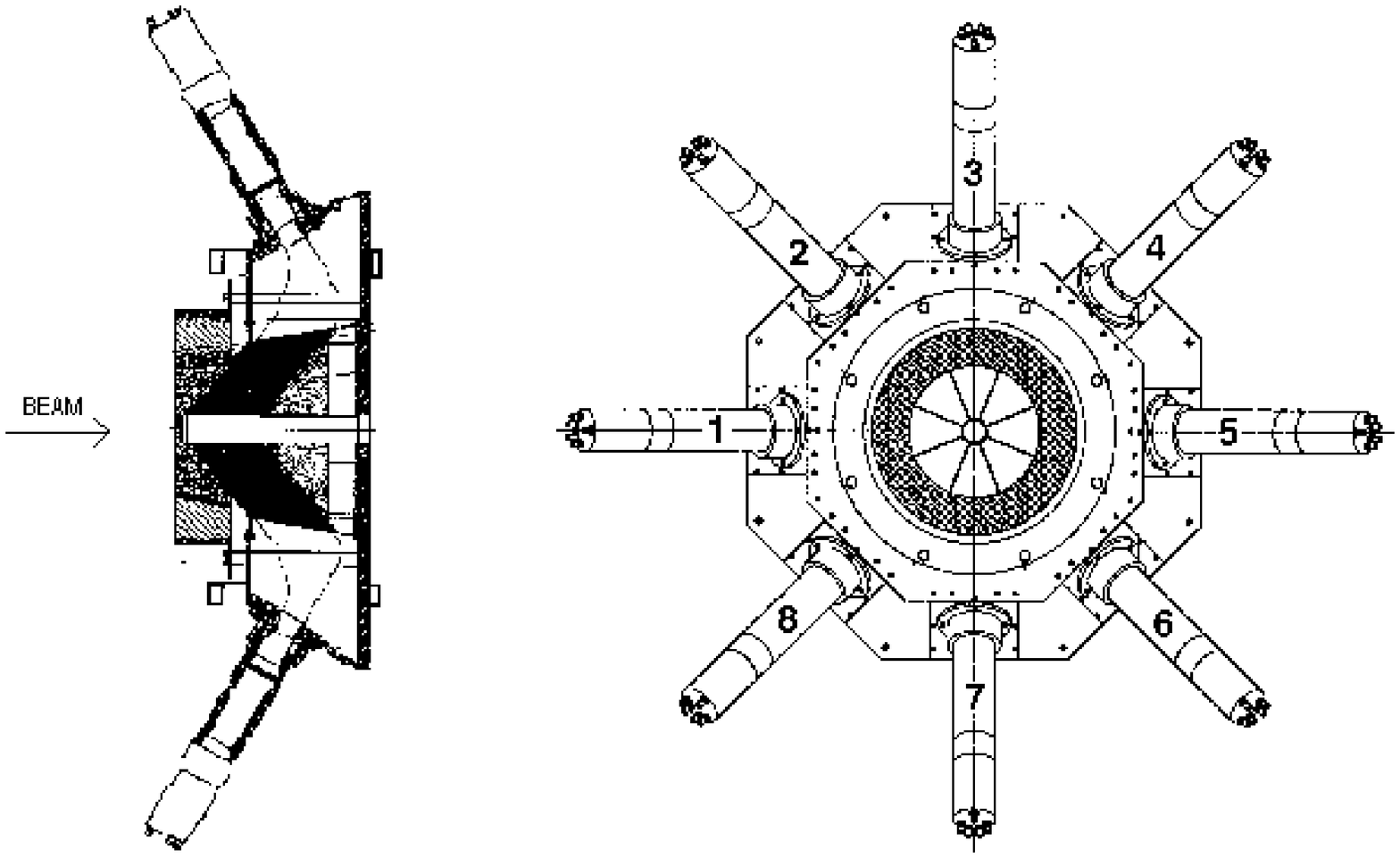}
}%
\caption{side and front schematic view of the Very Forward Electromagnetic calorimeter for the NA52 experiment.}
\label{fig:QC:qcal_na52}
\end{figure}

The NA52 or NEWMASS \cite{ref:QC:newmass,ref:QC:na52} was a fixed target 
experiment at the heavy ion physics program at CERN-SPS. Its objectives 
included detection of strangelets and study of anti-nuclei production in 
collisions of $Pb$ beam on $Pb$ target, at 158~GeV/ nucleon beam energy.
A very forward electromagnetic calorimeter \cite{ref:QC:weber97} was used to 
measure the energy of photons, produced from $\pi^0$ decays, in order to 
determine the centrality of the collision. It was placed at 0.8~m after the 
target, covering pseudorapidity interval $ 2.6 \leq \eta \leq 4.3$, and 
2$\pi$ in azimuth. The calorimeter was equipped with quartz fibers to cope 
with the high radiation dose expected and for fast response. It consisted 
of consecutive absorber-fibers layers (``sandwich'' geometry) inclined at
$45^{\circ}$ angle with respect to the beam axis. The calorimeter was also 
azimuthally segmented in 8 sectors to facilitate studies on anisotropy of 
transverse energy flow. The main features of the calorimeter are tabulated in 
table~\ref{tab:QC:qcal_na52}, a schematic view is given in 
fig.~\ref{fig:QC:qcal_na52}.

\begin{table}[p]
\centering
\caption{specifications of the Very Forward Electromagnetic calorimeter for the NA52 experiment.}
\label{tab:QC:qcal_na52}
\vspace{10pt}
{\footnotesize
\begin{tabular}{ll}
\hline\hline
                   & \hfill \ \\
{\bf  purpose}     & photon (from $\pi^0$ decay) energy measurement for impact parameter \\
                   & determination and for studies on transverse energy flow anisotropy\\
                   & \\
{\bf  position}    & 0.8~m after the target, covering pseudorapidity $ 2.6 \leq \eta \leq 4.3$ \\
                   & and 2$\pi$ in azimuth\\
                   & \\
{\bf  construction}& 19 consecutive absorber-fibers layers, \\
                   & at $45^{\circ}$ inclination with respect to beam axis, depth 19~$X_0$,\\
                   & azimuthal segmentation in 8 sectors,\\
                   & 1 photomultiplier per sector (total: 8)\\
                   & \\
{\bf  absorber}    & lead (Pb: $\lambda_{I}$= 17.2~cm, $X_0$= 0.56~cm, density= 11.3~gr/cm$^3$),\\
                   & 19 layers, 4~mm thick each  \\
                   & \\
{\bf  fibers}      & quartz core ($\varnothing$ 0.43~mm), hard plastic cladding,\\
                   & numerical aperture NA=0.22\\
                   & 1 fiber plane per absorber layer\\
                   & \\
{\bf filling ratio}& $\frac{\textrm{fiber volume}}{\textrm{absorber volume}}$ : 1/12.8 (= 7.8\%) \\
                   & \\
{\bf $\sigma$(E)/E}& $\frac{0.56}{\sqrt{E(GeV)}} \oplus 3.6\%$ \\
                   & \\
{\bf total dose}   & 1 Grad \\
                   & \\
\hline\hline
\end{tabular}
}
\end{table}

\subsection{The Zero Degree calorimeters of the experiments at BNL-RHIC}

\begin{figure}[p]
\centering
\rotatebox{-90}{%
\epsfxsize=180pt
\epsffile{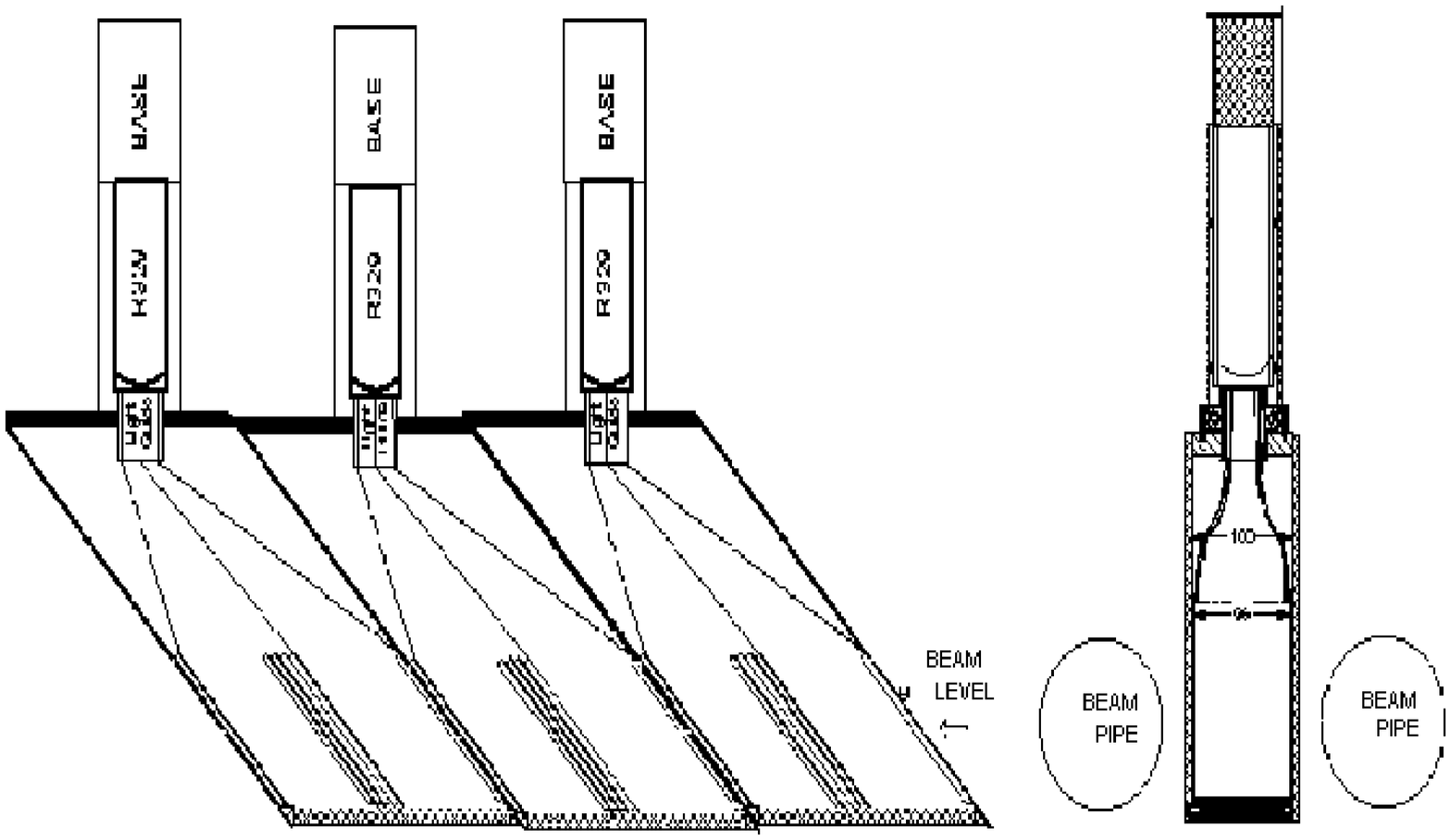}
}%
\caption{side and front view of the Zero Degree calorimeters for the RHIC experiments.}
\label{fig:QC:qcal_rhic}
\end{figure}

The Relativistic Heavy Ion Collider at BNL provides colliding $Au$ ion beams
at 100~GeV per nucleon ($\sqrt{S}_{Au+Au}= 39.4$~TeV), lighter ion beams at 
125~GeV/nucleon and proton beams at 250~GeV/proton. There are 4 experiments 
\cite{ref:QC:wwwrhic,ref:QC:rhic} dedicated to the detection and study of the 
quark-gluon plasma phase. Each experiment is equipped with a pair of identical 
Zero Degree calorimeters \cite{ref:QC:rhiczdc98,ref:QC:adler2000}, positioned 
symmetrically on both sides of the interaction point, and at 18~m distance. 
The calorimeters determine the event centrality and provide triggering signal 
by measuring the energy of the nucleons that do not participate in the 
collision. They also operate as luminosity monitors 
\cite{ref:QC:white98,ref:QC:baltz98}. Each calorimeter consists of consecutive 
absorber-fibers layers that are inclined $45^{\circ}$ with respect to the 
beam axis. The specifications of the calorimeters and their schematic views 
are given in table~\ref{tab:QC:qcal_rhic} and depicted in 
fig.~\ref{fig:QC:qcal_rhic}, respectively.

\begin{table}[p]
\centering
\caption{specifications of the Zero Degree calorimeters for the RHIC experiments.}
\label{tab:QC:qcal_rhic}
\vspace{10pt}
{\footnotesize
\begin{tabular}{ll}
\hline\hline
                   & \hfill \ \\
{\bf  purpose}     & measurement of spectator neutrons' energy for impact parameter determination \\
                   & and for luminosity monitoring \\
                   & \\
{\bf  position}    & $\pm$18~m from the interaction point, covering pseudorapidity $|\eta| \geq 6.7$ \\ 
                   & \\
{\bf  construction}& 81 consecutive absorber-fibers layers of $10\times18.7$~cm$^2$,\\
                   & at $45^{\circ}$ inclination with respect to beam axis, depth 5.1~$\lambda_{I}$,\\
                   & 1 photomultiplier per 27 layers (total: 3)\\
                   & \\
{\bf  absorber}    & tungsten alloy ($\lambda_{I}$= 11.2~cm, $X_0$= 0.38~cm),\\
                   & 81 layers 5~mm each  \\
                   & \\
{\bf  fibers}      & PMMA core ($\varnothing$ 0.45~mm), silica fluorinated cladding ($\varnothing$ 0.5~mm),\\
                   & numerical aperture NA=0.50 \\
                   & 1 fiber plane per absorber layer\\
                   & \\
{\bf filling ratio}& $\frac{\textrm{fiber volume}}{\textrm{absorber volume}}$ : 8\% \\
                   & \\
{\bf $\sigma$(E)/E}& $\frac{0.846}{\sqrt{E(GeV)}} + 9.1\%$ (= 17\% at 100~GeV) \\
                   & \\
{\bf light yield}  & 5.2 photoelectrons per GeV (protons)\\
                   & \\
{\bf total dose}   & 100 krad \\
                   & \\
\hline\hline
\end{tabular}
}
\end{table}

\subsection{The Very Forward EM calorimeter for the H1 experiment at DESY-HERA}

\begin{figure}[p]
\centering
\epsfysize=250pt
\epsffile{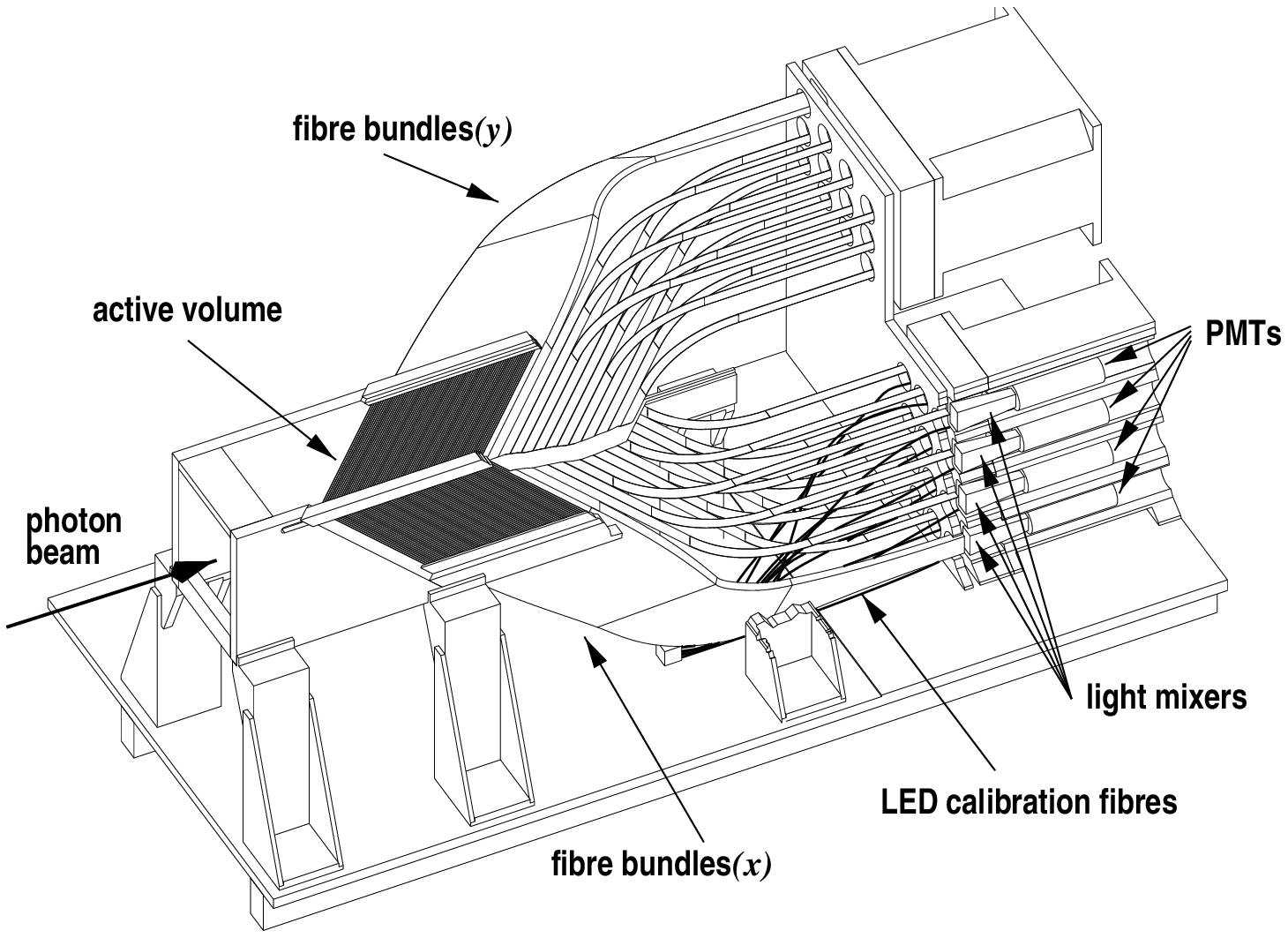}
\caption{general view of the Very Forward Electromagnetic calorimeter for the H1 experiment.}
\label{fig:QC:qcal_h1}
\end{figure}

The H1~\cite{ref:QC:experH1_1,ref:QC:experH1_2} is one of the experiments at 
the DESY-HERA $ep$ collider (E$_e$= 27.5 GeV, E$_p$= 920~GeV). Its goal is to 
study the structure of the proton and the properties of its constituents.
An electromagnetic calorimeter~\cite{ref:QC:andrieu2001},
which measures the energy of photons produced by bremsstrahlung of electrons
at the interaction point, is used as a luminosity monitor. The calorimeter has 
quartz fibers as sensitive material to withstand the radiation level. It consists of 
consecutive absorber-fibers trapezoid layers with $45^{\circ}$ inclination 
with respect to the beam. A special feature of this calorimeter is that it 
is also capable to measure the impact point of the incident photons.
In particular, this is achieved by alternating fiber planes with horizontal 
and vertical orientation. Each fiber plane is divided in 12 fiber strips of 
10~mm width. The calorimeter signal is read out by 12+12 photomultipliers,
for the horizontal and for the vertical set of strips, respectively.
With such a configuration, a position resolution of 
$\sigma$= 5~mm/$\sqrt{E(GeV)}$ is achieved. The construction parameters and 
the main properties of the calorimeter are tabulated in 
table~\ref{tab:QC:qcal_h1}. A general view is shown in 
fig.~\ref{fig:QC:qcal_h1}.

\begin{table}[p]
\centering
\caption{specifications of the Very Forward Electromagnetic calorimeter for the H1 experiment.}
\label{tab:QC:qcal_h1}
\vspace{10pt}
{\footnotesize
\begin{tabular}{ll}
\hline\hline
                   & \hfill  \\
{\bf  purpose}     & luminosity monitoring by measuring the energy of photons produced \\
                   & by bremsstrahlung of electrons at the interaction point \\
                   & \\
{\bf  position}    & $\sim$100~m from the interaction point\\
                   & \\
{\bf  construction}& 70 consecutive absorber-fibers layers, at $45^{\circ}$ inclination, depth 25~$X_0$,\\
                   & alternating fiber planes with horizontal and vertical orientation,\\
                   & a fiber plane is divided in 12 strips of 10~mm width each,\\
                   & 12+12 photomultipliers (for horizontal + vertical fiber strips) \\
                   & to determine the x, y impact point of the incident photon \\
                   & \\
{\bf  absorber}    & tungsten (W: $X_0$= 0.356~cm), 70 layers 0.7~mm thick each  \\
                   & \\
{\bf  fibers}      & quartz core ($\varnothing$ 0.60~mm), PMMA cladding ($\varnothing$ 0.64~mm), NA=0.37\\
                   & 1 fiber plane per absorber layer, \\
                   & a fiber plane is divided in 12 strips of 10~mm width each\\
                   & \\
{\bf filling ratio}& $\frac{\textrm{fiber volume}}{\textrm{absorber volume}}$ : 1/1.68 (= 59.5\%) \\
                   & \\
{\bf $\sigma$(E)/E}& $\frac{0.19}{\sqrt{E(GeV)}} \oplus 0.5\%$  \hspace{50pt}{\bf $\sigma$(position)} = $\frac{5~mm}{\sqrt{E(GeV)}}$\\
                   & \\
{\bf light yield}  & 130 photoelectrons per GeV (electrons)\\
                   & \\
{\bf total dose}   & 0.1 Grad \\
                   & \\
\hline\hline
\end{tabular}
}
\end{table}

\subsection{The CASTOR calorimeter for the CERN-LHC}

\begin{figure}[p]
\centering
\epsfysize=170pt
\epsffile{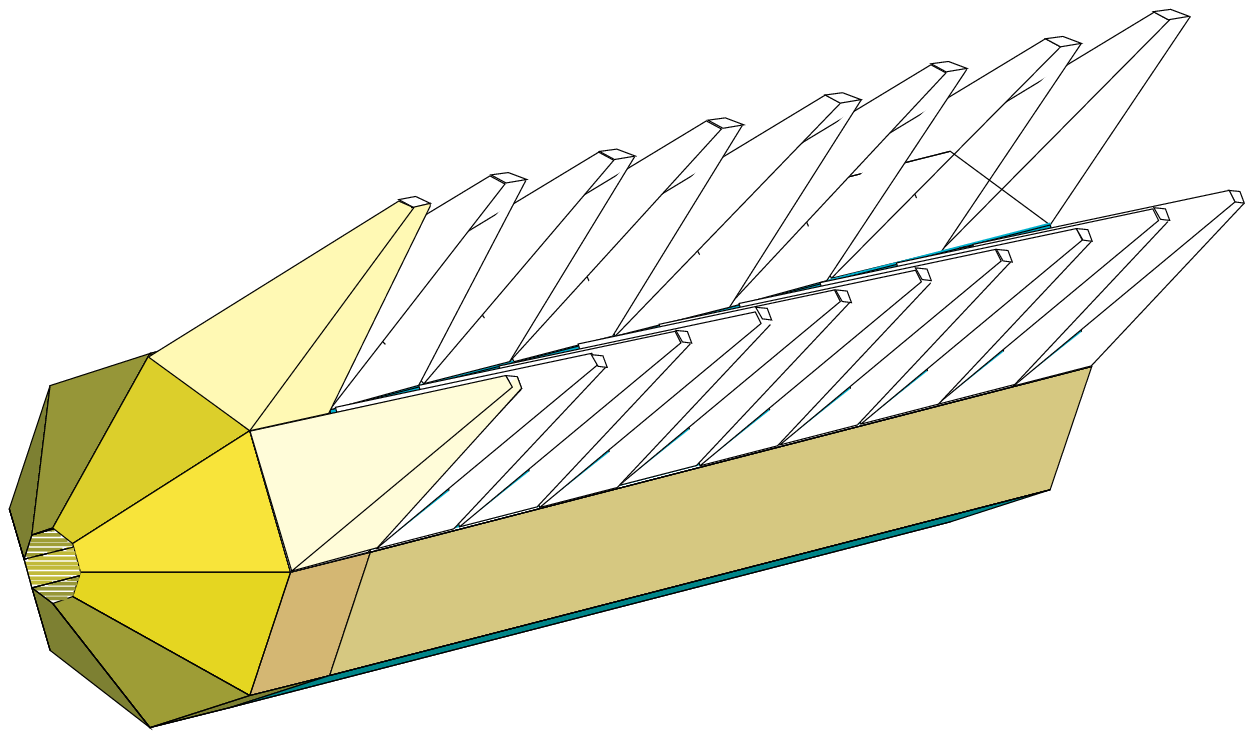}
\caption{schematic view of the CASTOR calorimeter, some of its aircore lightguides are shown.}
\label{fig:QC:castor}
\end{figure}

The calorimeter is proposed as a very forward detector of the ALICE or CMS
experiments at the LHC. Its main objective is to search in the baryon rich, 
very forward rapidity region of central $Pb+Pb$ collisions for unusual events, 
the so called Centauro events, and for ``long penetrating objects'', assumed 
to be strangelets, by measuring the hadronic and electromagnetic energies and 
the longitudinal profile of the hadronic showers 
\cite{ref:QC:castor02}-\cite{ref:QC:gm_phd}.
It is planned to cover the pseudorapidity range $5.46 \leq\eta\leq 7.14$,
and 2$\pi$ in azimuth. The calorimeter is azimuthally divided in 8 sectors 
and longitudinally segmented in consecutive layers of W absorber interleaved
with quartz fiber planes having 45$^{\circ}$ inclination with respect to the 
beam axis. The calorimeter is composed of a number of channels per sector, 
read-out by aircore lightguides that will collect and transmit the signal of 
each channel to its corresponding photomultiplier 
(see fig.~\ref{fig:QC:castor}) \cite{ref:QC:gm3,ref:QC:gm_phd}.
The specifications and the main properties of the calorimeter are tabulated 
in table~\ref{tab:QC:castor}. A general view is shown in
fig.~\ref{fig:QC:castor}.

\begin{table}[p]
\centering
\caption{specifications of the CASTOR calorimeter.}
\label{tab:QC:castor}
\vspace{10pt}
{\footnotesize
\begin{tabular}{ll}
\hline\hline
                   & \hfill  \\
{\bf  purpose}     & measurement of hadronic and electromagnetic energies and \\
                   & longitudinal profile of hadronic showers in central $Pb+Pb$ collisions \\
                   & \\
{\bf  position}    & 16.4~m from the interaction point, covering $5.46 \leq\eta\leq 7.14$\\
                   & in pseudorapidity and 2$\pi$ in azimuth\\
                   & \\
{\bf  construction}& azimuthal segmentation in 8 sectors, \\
                   & 230 consecutive absorber-fibers layers per sector \\
                   & at $45^{\circ}$ inclination with respect to beam axis, depth 10~$\lambda_{I}$ \\
                   & 1 lightguide+photomultiplier per 1 $\lambda_{I}$ per sector (total: 8 $\times$ 10)\\
                   & \\
{\bf  absorber}    & tungsten (W: $\lambda_{I}$= 10.0~cm, $X_0$= 0.365~cm, density= 18.5~gr/cm$^3$)\\
                   & 230 layers per sector, 0.3 cm thick each ($d/X_0 = 1.16$)\\
                   & \\
{\bf  fibers}      & quartz core ($\varnothing$ 0.60~mm), hard plastic cladding ($\varnothing$ 0.64~mm), \\
                   & numerical aperture NA=0.37, \\
                   & 2 fiber planes per absorber layer\\
                   & \\
{\bf filling ratio}& $\frac{\textrm{fiber volume}}{\textrm{absorber volume}}$ : 29.5\% \\
                   & \\
{\bf $\sigma$(E)/E}& $\frac{(21.\pm 0.3)\%}{\sqrt{E(GeV)}} \oplus (0.00\pm 0.04)\%$ for $e^{\pm}$, $\frac{(95.\pm 2.)\%}{\sqrt{E(GeV)}} \oplus (6.5\pm 0.3)\%$ for $\pi^{\pm}$\\
                   & \\
{\bf light yield}  & $\sim$40(30) photoelectrons per GeV for $e^{\pm}$($\pi^{\pm}$)\\
                   & \\
{\bf total dose}   & $\sim$300 krad \\
                   & \\
\hline\hline
\end{tabular}
}
\end{table}

\subsection{Other calorimeters}

To the knowledge of the author in order to complete the above list we 
should also mention the W/quartz fiber polarimeter at the SLD experiment 
at SLAC and the dual readout calorimeter R\&D project. 
The former is a calorimeter consisting of sandwiches of tungsten plates 
and fiber layers with  alternating vertical and horizontal orientation to 
allow position measurement. With a fiducial volume of 
$43.5 \times 43.5 \times 115$~mm$^3$ it was capable to measure the energy 
and the impact point of the Compton scattered photons in order to determine 
precisely the polarization of the electron beam 
\cite{ref:QC:berridge97,ref:QC:onoprienko2000}.
In the latter project, a hadronic calorimeter is proposed to be equipped with 
both scintillating and quartz fibers. A hadron shower generates signals in both 
materials but by different parts of the shower, as discussed in 
\ref{subsec:QC:principle_of_operation} and \ref{subsec:QC:basic_properties}. 
By measuring both signals simultaneously and combining them it is possible to 
determine the electromagnetic content of hadron showers on an event-by-event 
basis. This complementary information is then used to correct the energy 
reconstruction to achieve better response linearity and improve the energy 
resolution for hadronic showers. First results on this concept with a 
``spaghetti'' geometry prototype can be found in \cite{ref:QC:wigmans2004}.

\begin{table}[p]
\caption{summary table of the main parameters of the various quartz fiber calorimeters.}
\label{tab:QC:all_qcals}
\vspace{1pt}
\centering
\rotatebox{90}{%
\small{
\begin{tabular}{l|ccccccc}
\hline\hline
                       & \\
{\bf calorimeter}      & {\bf geometry}& {\bf absorber}   & {\bf depth}       & {\bf filling ratio}& {\bf $\sigma$(E)/E}                              & {\bf light yield}     & {\bf dose} \\
                       & \\
\hline
                       & \\
NA50 ZDC               & $0^{\circ}$      & tantalum         & 5.6~$\lambda_{I}$ & 5.88\%            & $\frac{2.9}{\sqrt{E(GeV)}} \oplus 19\%$   & 0.5 p.e./GeV  & 10 Grad \\
                       & \\
CMS VFCal              & $0^{\circ}$      & steel            & 8.8~$\lambda_{I}$ & $<$ 1\%           & $\frac{2.7}{\sqrt{E(GeV)}} \oplus 13\%$   & 0.5 p.e./GeV  & 1 Grad \\
                       & \\
ALICE nZDC             & $0^{\circ}$      & tantalum         & 8.5~$\lambda_{I}$ & 4.55\%            & 10.5\% ($n$ 2.7~TeV)                 & 0.35 p.e./GeV & $\cal O$(1) Grad \\
ALICE pZDC             & $0^{\circ}$      & brass            & 8.4~$\lambda_{I}$ & 1.54\%            & $\sim$10\% ($p$ 2.7~TeV)             & 0.28 p.e./GeV & $\cal O$(1) Grad \\
                       & \\
NA52 VFEMcal           & $45^{\circ}$     & lead             & 19~$X_0$          & 7.8\%             & $\frac{0.56}{\sqrt{E(GeV)}} \oplus 3.6\%$ &                    & 1 Grad \\
                       & \\
RHIC ZDC               & $45^{\circ}$     & tungsten         & 5.1~$\lambda_{I}$ & 8\%               & $\frac{0.846}{\sqrt{E(GeV)}} + 9.1\%$     & 5.2 p.e./GeV  & 100 krad \\
                       & \\
H1 VFEMcal             & $45^{\circ}$     & tungsten         & 25~$X_0$          & 59.5\%            & $\frac{0.19}{\sqrt{E(GeV)}} \oplus 0.5\%$ & 130 p.e./GeV  & 0.1 Grad \\
                       & \\
CASTOR                 & $45^{\circ}$     & tungsten         & 10~$\lambda_{I}$  & 29.5\%            & $\frac{0.21}{\sqrt{E(GeV)}} \oplus 0.0\%$ for $e^{\pm}$    & 40 p.e./GeV for $e^{\pm}$   & 300 krad \\
                       &                  &                  &                   &                   & $\frac{0.95}{\sqrt{E(GeV)}} \oplus 6.5\%$ for $\pi^{\pm}$  & 30 p.e./GeV for $\pi^{\pm}$ &          \\
                       & \\
\hline\hline
\end{tabular}
}
}
\end{table}

\section{Summary}

The quartz fiber calorimetry is a technique the signal generation mechanism 
of which is based on the Cherenkov effect. Namely, the charged particles of 
the shower with $\beta > \beta_{threshold}$ when traversing the fibers produce 
Cherenkov light. The light is captured and lightguided inside the fibers
and finally collected by photomultipliers. We described in detail the 
operation principle and the main properties of this technique and we discussed 
the quartz fiber calorimeters that have been built or planned to in various 
experiments. In summary, the advantages of quartz fiber calorimeters are,
the radiation hardness, the fast response and the compact detector dimensions. 
They operate as centrality detectors, trigger detectors, luminosity monitors or 
general purpose calorimeters at very forward regions.

\subsection*{Acknowledgements}

Many thanks to N.~Saoulidou for valuable discussions and comments.

\clearpage

\addcontentsline{toc}{section}{References}

\columnsep 10pt
\begin{multicols}{2}

\end{multicols}

\end{document}